\documentclass[prd,tightenlines,showpacs,nofootinbib,eqsecnum,amsfonts,amssymb,amsmath,twocolumn,widetext,preprintnumbers]{revtex4}
\usepackage{graphicx}
\usepackage{amsmath}
\usepackage{amsfonts}   
\usepackage{amssymb}  
\usepackage{mathrsfs}
\usepackage{epsfig}
\usepackage{bm}
\usepackage{color}
\begin{document}
\newcommand{\average}[1]{\langle{#1}\rangle_{{\cal D}}}
\newcommand{\dd}{{\rm d}}
\newcommand{\etal}{{\it et al.}}
\preprint{IPMU13-0008}

\title{From configuration to dynamics\\ {\it Emergence of Lorentz signature in classical field theory}}

\author{Shinji Mukohyama}
\email{shinji.mukohyama@ipmu.jp}
 \affiliation{
             Kavli Institute for the Physics and Mathematics of the
	     Universe, Todai Institutes for Advanced Study, University
	     of Tokyo, 
             5-1-5 Kashiwanoha, Kashiwa, Chiba 277-8583, Japan}
             
\author{Jean-Philippe Uzan}
\email{uzan@iap.fr}
 \affiliation{
             Institut d'Astrophysique de Paris,
             Universit\'e Pierre~\&~Marie Curie - Paris VI,
             CNRS-UMR 7095, 98 bis, Bd Arago, 75014 Paris, France.}

\begin{abstract}
 The Lorentzian metric structure used in any field theory allows one to
 implement the relativistic notion of causality and to define a notion
 of time dimension. This article investigates the possibility that at
 the microscopic level the metric is Riemannian, i.e. locally Euclidean, 
 and that the Lorentzian structure, that we usually consider as
 fundamental, is in fact an effective property that emerges in some
 regions of a 4-dimensional space with a positive definite metric. In
 such a model, there is no dynamics nor signature flip across some
 hypersurface; instead, all the fields develop a Lorentzian dynamics in
 these regions because they propagate in an effective metric. It is
 shown that one can construct a decent classical field theory for
 scalars, vectors and (Dirac) spinors in flat spacetime. It is then
 shown that gravity can be included but that the theory for the
 effective Lorentzian metric is not general relativity but of the
 covariant Galileon type. The constraints arising from stability, the
 equivalence principle and the constancy of fundamental constants are
 detailed and a phenomenological picture of the emergence of the
 Lorentzian metric is also given. The construction, while restricted to
 classical fields in this article, offers a new view on the notion of
 time.
\end{abstract}
\date{\today}
\maketitle
\section{Introduction}\label{sec1}

When constructing a physical theory, there is a large freedom in the
choice of the mathematical structures. The developments of theoretical
physics taught us that some of these structures are well-suited to
describe some classes of phenomena (e.g. the use of a vector field for 
electromagnetism, of spinors for some class of particles, the use of some
symmetries, etc.). However, these choices can only be validated by the
mathematical consistency of the theory and the agreement between the
consequences of these structures and experiments. It may even be that 
different structures are possible to reproduce what we know about
physics and one may choose one over the other on the basis of less
well-defined criteria such as simplicity and economy. 

At each step, some properties such as the topology of
space~\cite{topology}, the number of spatial dimensions or the numerical
values of the free parameters that are the fundamental
constants~\cite{jpucte}, may remain a priori free in a given framework,
or imposed in another framework (e.g. the number of space dimensions is
fixed in string theory~\cite{stringT}). 

Among all these structures, and in the framework of metric theories of
gravitation, the signature of the metric is in principle
arbitrary. Indeed, it seems that on the scales that have been probed so
far there is the need for only one time dimension and three spatial
dimensions. In special and general relativity, time and space are
geometrically different because the geometry of spacetime is locally
Minkowskian, i.e. it enjoys a Lorentzian metric with signature
$(-,+,+,+)$, i.e. the line element is 
$\dd s^2=-\dd t^2 + \dd {\bm\ell}^2\equiv\eta_{\mu\nu}\dd x^\mu \dd x^\nu$. 
While the existence of two time directions may lead to
confusion~\cite{V4}, it is not clear if there are theoretical obstacles
to have more than one time directions, as even suggested by some
framework~\cite{V5} (see the argument for such possibilities made by
Ref.~\cite{no2} and detailed further in Ref.~\cite{gibbons}). Several
models for the birth of the universe~\cite{nothing} are based on a
change of signature via an instanton in which a Riemannian and a
Lorentzian manifolds are joined across a hypersurface which may be
thought of as the origin of time. While there is no time in the
Euclidean region, where the signature is $(+,+,+,+)$, it flips to
$(-,+,+,+)$. Eddington even suggested~\cite{G85} that it can flip
across some surface to $(-,-,+,+)$. Signature flip also arises in
brane-world scenarios~\cite{P86} (see Ref.~\cite{gibbons} for a review
of these possibilities) or in loop quantum cosmology~\cite{lqc}. These 
discussions however let the problem of the origin of the time direction
open~\cite{zeh}. 

In Newtonian theory, time is a fundamental concept. It is assumed to
flow and is described by a real variable. It can be measured by good
clocks and any observers shall, irrespective of their motion, agree on 
the time elapsed between two events~\cite{vuibert}. The laws of dynamics 
describe the change of configurations of a system with time. In  
relativity, first the notions of space and time are set on the same
footing and second, the notion of time is no more unique. One has to
distinguish between a coordinate time, with no physical meaning, and the
proper time that can be measured by an observer. Quantum mechanics
offers another insight on time: there, while there may be operators or
observables corresponding to spatial positions, time is not an
observable, and thus not an operator~\cite{rovelli0}. As detailed in
Ref.~\cite{gibbons}, by an argument going back to Pauli, commutation
relations like $[\hat x^\mu,\hat P_\nu]=i\delta^\mu_\nu$ are
incompatible with the spectrum of $\hat P^\mu$ lying in the future
lightcone and the notion of time is intimately related to the complex
(Hilbert Space) structure of quantum mechanics~\cite{gibbons}.

The question of whether time does actually ``exist'' has been widely
debated in the context of classical physics~\cite{barbour},
relativity~\cite{time-RG} and quantum mechanics~\cite{rovelli0}. The
debate on the nature of time has shifted with quantum gravity where the
recovery of a classical notion of time is considered as a problem. In
that case, the Schr\"odinger equation becomes the Wheeler-de Witt
equation, of the form $\hat H\vert\Psi\rangle=0$, so that the allowed
states are those for which the Hamiltonian vanishes. Thus, it determines
in which states the universe can be but does not give any evolution
through time. We refer to Refs.~\cite{rovelli,time,Ellis:2006sq} for general
discussions on the nature of time. This has led to numerous works on the
emergence of time in different versions of quantum 
gravity~\cite{rovelli2,markopoulou,time-emergence,time-emergence2,time-emergence3,time-emergence0}
(and indeed the reverse opinion has been argued~\cite{carroll}). Also, 
the thermodynamical aspects of gravity, the existence of dualities
between gauge theories and gravity theories~\cite{maldacena}, and
holography~\cite{holo} have led to the idea that the metric itself may 
have to be thought of as the result of a coarse-graining of underlying
more fundamental degrees of freedom~\cite{V6}.  

The local Minkowski structure is an efficient way to implement the
notion of causality in realistic theories and is today accepted as a
central ingredient of the construction of the relativistic theory of
fields. When gravity is included, the equivalence principle implies
(this is not a theoretical requirement, but just an experimental fact,
required at a given accuracy) that all the fields are universally
coupled to the same Lorentzian metric. From the previous discussion, we
may wonder whether the signature of this metric is only a convenient way
to implement causality or whether it is just a property of an effective
description of a microscopic theory in which there is no such notion. 

This article proposes the view according to which the fundamental
physical theory is intrinsically purely Euclidean so that its field
equations determine a static 4-dimensional field configuration. The
Lorentzian dynamics that we can observe in our universe has then to be
thought of as an emergent property, that is as an illusion holding in a
small patch of a Euclidean mathematical space. This is thus an attempt to
go further than early proposals~\cite{Greensite:1992np,Girelli,Goulart} and see
to which extent this can be an open possibility. We emphasize that it is
different from the models discussed above involving a signature change
across a boundary or obtained by rotating to an Euclidean space. We
consider it important to take the freedom to see how far one can go in
such a direction. As we shall later discuss, if possible, such a setting
may shed a new light on several theoretical issues from the nature of
singularities to quantum gravity.

Our attitude is however more modest and we want to start by constructing
a decent classical field theory under this hypothesis. 
Section~\ref{sec2} explains the basics of our mechanism and then
describes the construction of the scalar, vector and spinor sectors in
flat spacetime. We show that the whole standard model of particle
physics can be constructed from a Euclidean theory, at the classical
level. Section~\ref{sec3} addresses the more difficult question of
gravity. While general relativity is not recovered in general, it shows
that an extended $K$-essence theory of gravity called covariant Galileon
can be obtained. We then show in Section~\ref{sec4} that the dynamics of
scalar and vector in curved spacetime can also be obtained. We then
discuss the experimental and theoretical constraint on our construction
in \S~\ref{sec5} and also propose a way to understand phenomenologically
the emergence of the effective Lorentzian dynamics. It is however to be
remembered that there is no dynamics at the fundamental level and that
this illusion is restricted to a domain of a large Euclidean space.

\section{Field theory in flat space}\label{sec2}

This section introduces the mechanism in the simple case of a flat space
(\S~\ref{subsec2.0}). It shows how scalars  (\S~\ref{subsec2.1}),
vectors  (\S~\ref{subsec2.2} and  \S~\ref{subsec2.3}) and spinors
(\S~\ref{subsec2.4}) defined in Euclidean space can have an apparent
Lorentzian dynamics. We finish by pointing out the properties and limits
of this mechanism in \S~\ref{subsec2.5}, many of them being discussed in
a more realistic version in the following sections. 

\subsection{Clock field}\label{subsec2.0}

\begin{figure}[h!]
\centering
\includegraphics[width=\columnwidth]{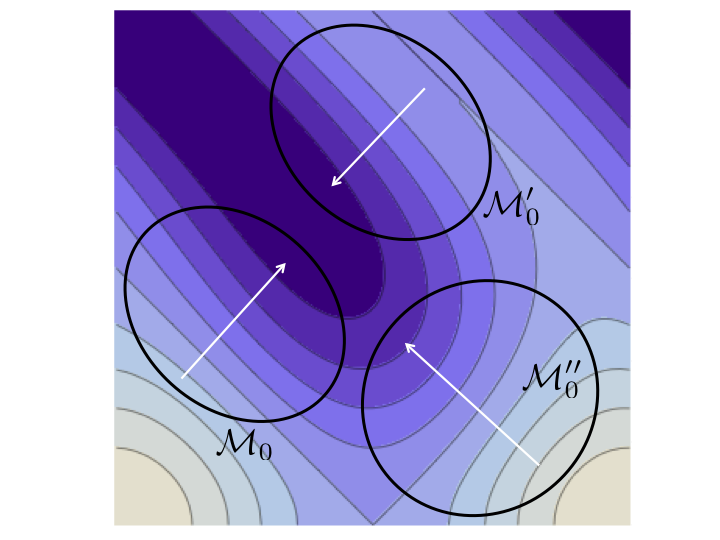}
 \caption{Example of a spatial configuration of the clock field. Locally,
 one can define regions such as ${\cal M}_0$, ${\cal M}'_0$ and 
${\cal M}''_0$, in each of which a time direction emerges. Indeed this
 direction does not preexist at the microscopic level and can be
 different from patches to patches.} 
\label{fig1}
\end{figure}

In order to understand the basics of our model, let us consider a
$4$-dimensional Riemannian manifold ${\cal M}$ with a positive definite
Euclidean metric $g^{\rm E}_{\mu\nu}=\delta_{\mu\nu}$  in a Cartesian 
coordinate system. As a consequence, the theory we shall consider on
this manifold does not have a natural concept of time. In order to make
such a notion emerge locally, we introduce a scalar field $\phi$ and
assume that its derivative has a non-vanishing vacuum expectation value
(vev) in a region ${\cal M}_0$ of the Riemannian space (see
Fig.~\ref{fig1}). To be more precise, we assume that
$\partial_\mu\phi=\hbox{const.}\not=0$ in ${\cal M}_0$. It follows that
we can set 
\begin{equation}
 \partial_\mu\phi= M^2 n_\mu \quad\hbox{in}\quad {\cal M}_0
\end{equation}
with $n_\mu$ a unit constant vector
($\delta^{\mu\nu}n_{\mu}n_{\nu}=1$). We have introduced a mass 
scale $M$ so that $n_\mu$ is dimensionless. By construction, its norm
$X_{\rm E}\equiv\delta^{\mu\nu}\partial_{\mu}\phi\partial_{\nu}\phi=M^4$
is constant and satisfies 
\begin{equation}
 X_{\rm E}>0 \quad\hbox{in}\quad {\cal M}_0.
\end{equation}
Now, under this assumption, one of the coordinates  can be chosen as
\begin{equation}
 \dd t = n_\mu \dd x^\mu.
\end{equation}
This accounts for choosing
\begin{equation}\label{e.time0}
 t\equiv \frac{\phi}{M^2}
\end{equation}
up to a constant that can be set to zero without loss of generality. The
metric of the 4-dimensional Riemannian space (with Euclidean geometry)
can be rewritten as 
\begin{eqnarray}
 \dd s_{\rm E}^2&=&\delta_{\mu\nu}\dd x^\mu\dd x^\nu\nonumber\\
                    &=&\left(n_\mu \dd x^\mu\right)^2 + \left(\delta_{\mu\nu}-n_\mu n_\nu\right)\dd x^\mu\dd x^\nu\nonumber\\
                    &=& \dd t^2 + \delta_{ij}\dd x^i\dd x^j,
\end{eqnarray}
by introducing a set of three independent coordinates $x^i$
($i=1\ldots3$) on the hypersurfaces $\Sigma_t$ normal to $n^\mu$. Note
that the geometry on $\Sigma_t$ would not be Euclidean if $n_\mu$ were
not constant. As we shall now discuss, the scalar field $\phi$ will be
related to what we usually call ``time'', so that we shall call such a
scalar field a {\it clock field}.

\subsection{Scalar field}\label{subsec2.1}

The Euclidean configuration of a scalar field $\chi$ can be obtained by
combining the usual action for a scalar field, with a kinetic term and a
potential, 
$$
-\int \dd^4x\left[ \frac{1}{2}\delta^{\mu\nu}\partial_\mu\chi\partial_\nu\chi + V(\chi)\right],
$$
with a coupling  to the clock field $\phi$ as
$$
 \int \dd^4x \left( \delta^{\mu\nu}\partial_\mu\phi\partial_\nu\chi\right)^2.
$$
Let us consider  the action obtained by the following combination
\begin{eqnarray}\label{e.chi1}
 S_\chi &=&  \int \dd^4x \left[ - \frac{1}{2}\delta^{\mu\nu}\partial_\mu\chi\partial_\nu\chi   -V(\chi)\right.\nonumber\\
 &&\left.\qquad\qquad+\frac{1}{M^4}\left( \delta^{\mu\nu}\partial_\mu\phi\partial_\nu\chi\right)^2
 \right].
\end{eqnarray}
It is straightforward to conclude that since $\delta^{\mu\nu}\partial_\mu\chi\partial_\nu\chi =(\partial_t\chi)^2+\delta^{ij}\partial_i\chi\partial_j\chi$ and, when restricted to ${\cal M}_0$,  $(\delta^{\mu\nu}\partial_\mu\phi\partial_\nu\chi)^2=M^4(\partial_t\chi)^2$, the action~(\ref{e.chi1}) reduces to
\begin{eqnarray}
 S_\chi &=&  \int \dd t\dd^3x \left[\frac{1}{2}\left(\partial_t\chi\right)^2 - \frac{1}{2}\delta^{ij}\partial_i\chi\partial_j\chi   -V \right]
\end{eqnarray}
in ${\cal M}_0$. This can indeed be rewritten as
\begin{eqnarray}
 S_\chi &=&  \int \dd t\dd^3x \left[ - \frac{1}{2}\eta^{\mu\nu}\partial_\mu\chi\partial_\nu\chi   -V \right].
\end{eqnarray}
The action~(\ref{e.chi1}) thus describes, when restricted to ${\cal M}_0$, the dynamics of a scalar field propagating in a 4-dimensional Minkowski spacetime with metric $\eta_{\mu\nu}=\hbox{diag}(-1,+1,+1,+1)$. The apparent Lorentzian dynamics, with a preferred time direction, is thus the result of the coupling to the scalar clock field.

\subsection{Vector field}\label{subsec2.2}

Usually, the dynamics of a vector field $A_\mu$  is dictated by the action $F_{\mu\nu}F^{\mu\nu}_{\rm E}$ where $F_{\mu\nu}$ is the Faraday tensor defined as $F_{\mu\nu}=\partial_\mu A_\nu-\partial_\nu A_\mu$ and where the subscript $E$ indicates that the indices are raised with the Euclidean metric $\delta^{\mu\nu}$.
 
The standard action of the vector field can be extended to include a coupling to the clock field of the form $F^{\mu\rho}_{\rm E}F^\nu_{E\rho}\partial_\mu\phi\partial_\nu\phi$ so that the action for the vector field we consider is
\begin{eqnarray}\label{e.A1}
 S_A = \frac{1}{4}\int\dd^4 x\left[ -F_{\mu\nu}F^{\mu\nu}_{\rm E}+\frac{4}{M^4}F^{\mu\rho}_{\rm E}F^\nu_{\rm E\rho}\partial_\mu\phi\partial_\nu\phi\right].
\end{eqnarray}
Since $F_{\mu\nu}F^{\mu\nu}_{\rm E}=2\delta^{ij}F_{0i}F_{0j}+\delta^{ik}\delta^{jl}F_{ij}F_{kl}$ and
since $F^{\mu\rho}_{\rm E}F^\nu_{{\rm E}\rho}\partial_\mu\phi\partial_\nu\phi=M^4\delta^{ij}F_{0i}F_{0j}$ in ${\cal M}_0$, it is easily concluded that this action can be rewritten as
\begin{equation}
 S_A=\frac{1}{4}\int\dd t\dd^3x\left[2\delta^{ij}F_{0i}F_{0j}-\delta^{ik}\delta^{jl}F_{ij}F_{kl} \right],
\end{equation}
or more simply as
\begin{equation}
 S_A=-\frac{1}{4}\int\dd t\dd^3x \,\eta^{\mu\alpha}\eta^{\nu\beta}F_{\mu\alpha}F_{\nu\beta}.
\end{equation}
Because of the coupling of the Faraday tensor to the clock field in the
Euclidean theory, the vector field propagates effectively in a Minkowski
metric and we recover the standard Maxwell action for a vector
field. The generalization to an non-Abelian group is straightforward. 

\subsection{Charged scalar field}\label{subsec2.3}

The construction of \S~\ref{subsec2.1} can easily be generalized to a
complex scalar field charged under a $U(1)$. Considering a complex
scalar field $\omega$, we add to the standard kinetic term
$\delta^{\mu\nu}(D_\mu\omega)^*(D_\nu\omega)$ a coupling to the clock
field of the form 
$\delta^{\mu\nu}\vert\partial_\mu\phi D_\nu\omega\vert^2$, where
$D_\mu\equiv\partial_\mu -i q A_\mu$. The Euclidean action is then
chosen to be 
\begin{eqnarray}\label{e.omega1}
 S_\omega &=&\int\dd^4x \left[-\frac{1}{2}\delta^{\mu\nu}(D_\mu\omega)^*(D_\nu\omega) -U(\vert\omega\vert^2)\right.\nonumber\\
  &&\qquad\qquad \left.+ \frac{1}{M^4}\delta^{\mu\nu}\vert\partial_\mu\phi D_\nu\omega\vert^2 \right].
\end{eqnarray}
Following the same arguments as for the real scalar field $\chi$, this
action takes the form 
\begin{eqnarray}
 S_\omega &=&  \int \dd t\dd^3x \left[ - \frac{1}{2}\eta^{\mu\nu}(D_\mu\omega)^*(D_\nu\omega)   -U \right].
\end{eqnarray}
Again, the coupling to the clock field implies that the Euclidean
dynamics leads to an effectively Minkowskian dynamics for $\omega$. 

\subsection{Spinor fields}\label{subsec2.4}

The next step is to include fermions in such a way that the standard
Dirac dynamics emerges from an Euclidean action. Let us start by
comparing the standard Dirac algebra in Minkowski spacetime
(\S~\ref{subsubsec2.4.1}) and that in Euclidean space
(\S~\ref{subsubsec2.4.2}) before we propose a choice of Euclidean 
action for the fermions (\S~\ref{subsubsec2.4.3}). 

\subsubsection{Dirac matrices in Minkowski spacetime}\label{subsubsec2.4.1}

In a Minkowski spacetime with signature ($-+++$), Dirac matrices are $4\times 4$ matrices satisfying the anti-commutation relation 
\begin{equation}\label{e.Dmat}
 \{\gamma^{\mu},\gamma^{\nu}\} = -2\eta^{\mu\nu}.
\end{equation}
For concreteness, throughout this section we shall adopt the following form of the Dirac matrices in Minkowski spacetime
\begin{eqnarray}
 \gamma^0 & = & {\bm \sigma}_0\otimes {\bm\sigma}_1 = 
  \left(\begin{array}{cc}
   {\bm0} & {\bm\sigma}_0\\
	 {\bm\sigma}_0 &  {\bm0}
	  \end{array} \right), \nonumber\\
 \gamma^i & = & i{\bm\sigma}_i\otimes {\bm\sigma}_2 = 
  \left(\begin{array}{cc}
    {\bm0} & {\bm\sigma}_i\\
	 -{\bm\sigma}_i &   {\bm0}
	  \end{array} \right),
\end{eqnarray} 
where ${\bm \sigma}_0$ is the $2\times 2$ unit matrix and  ${\bm \sigma}_i$ ($i=1,2,3$) are Pauli matrices, 
\begin{equation}
 {\bm\sigma}_1 = 
  \left(\begin{array}{cc}
   0 & 1 \\
	 1 & 0
	\end{array} \right), \quad
 {\bm\sigma}_2 = 
  \left(\begin{array}{cc}
   0 & -i \\
	 i & 0
	\end{array} \right), \quad
 {\bm\sigma}_3 = 
  \left(\begin{array}{cc}
   1 & 0 \\
	 0 & -1
	\end{array} \right). 
\end{equation}
While $\gamma^0$ is Hermitian, $\gamma^i$ are anti-Hermitian. One then defines $\gamma^5$ by 
\begin{equation}
 \gamma^5 \equiv -i\gamma^0\gamma^1\gamma^2\gamma^3
  = {\bm\sigma}_0\otimes {\bm\sigma}_3
  = \left(\begin{array}{cc}
   {\bm\sigma}_0 &  0\\
	  0 & -{\bm\sigma}_0
	   \end{array} \right),
\end{equation}
which satisfies 
\begin{equation}
  (\gamma^5)^2 = {\bf 1}, \quad
 \{\gamma^5,\gamma^{\mu}\} = 0 \quad (\mu=0,\cdots, 3). 
\end{equation}
The matrices 
\begin{equation}
 S^{\mu\nu} \equiv \frac{i}{4}[\gamma^{\mu},\gamma^{\nu}]
\end{equation}
satisfy the algebra of Lorentz generators
\begin{equation}
 [S^{\mu\nu},S^{\rho\sigma}] = 
  i(\eta^{\nu\rho}S^{\mu\sigma}-\eta^{\mu\rho}S^{\nu\sigma}
  -\eta^{\nu\sigma}S^{\mu\rho}+\eta^{\mu\sigma}S^{\nu\rho}).
\end{equation}
Hence, the Lorentz transformation for a Dirac field $\psi$ is 
\begin{equation}
 \psi \to \Lambda_{\frac12}\psi, \quad
  \Lambda_{\frac12} = 
  \exp\left[-\frac{i}{2}\omega_{\mu\nu}S^{\mu\nu}\right],
\end{equation}
where $\omega_{\mu\nu}$ are real numbers. Concretely,
\begin{eqnarray}
 S^{0i}&=& -\frac{i}{2}
  \left(\begin{array}{cc}
   {\bm \sigma}^i & {\bm 0}\\
	 {\bm 0} & -{\bm \sigma}^i
	\end{array} \right), \nonumber\\
 S^{ij}&=& \frac{1}{2}\sum_{k=1}^3\epsilon^{ijk}
  \left(\begin{array}{cc}
   {\bm \sigma}^k & {\bm 0}\\
	 {\bm 0} & {\bm \sigma}^k
	\end{array} \right). 
\end{eqnarray}
While $S^{ij}$  are Hermitian, $S^{0i}$ are anti-Hermitian. As a consequence, $\Lambda_{\frac12}$ is not unitary in general. In particular this means that 
\begin{equation}
 \psi^{\dagger} \to \psi^{\dagger}\Lambda_{\frac12}^{\dagger}
  \ne  \psi^{\dagger}\Lambda_{\frac12}^{-1} 
\end{equation}
and that $\psi^{\dagger}\psi$ is not a scalar under Lorentz transformation.  However, it is easy to check that 
\begin{equation}
 \bar{\psi} \to \bar{\psi}\Lambda_{\frac12}^{-1}, \quad
  \bar{\psi} \equiv \psi^{\dagger}\gamma^0
\end{equation}
so that $\bar{\psi}\psi$ is a scalar under Lorentz
transformations. This is the reason why the Dirac action in Minkowski
spacetime is usually constructed as 
\begin{equation}\label{e.diracstd}
  S^{\rm M}_\psi =
   \int\dd^4x\, \bar\psi \left( i\gamma^\mu\partial_\mu -m\right)\psi.
\end{equation}

\subsubsection{$\gamma$ matrices in Euclidean space}\label{subsubsec2.4.2}

In a $4$-dimensional Euclidean space with  metric $\delta_{\mu\nu}$, one can also define matrices $\gamma_{\rm E}^{\mu}$ according to
\begin{equation}
 \gamma_{\rm E}^0 \equiv i\gamma^5, \quad
  \gamma_{\rm E}^i \equiv \gamma^i
\end{equation} 
so that they obey the anti-commutation relation
\begin{equation}
 \{\gamma_{\rm E}^{\mu},\gamma_{\rm E}^{\nu}\} = -2\delta^{\mu\nu}.
\end{equation} 
Then, we can define 
\begin{equation}
 \gamma_{\rm E}^5 \equiv \gamma_{\rm E}^0\gamma_{\rm E}^1\gamma_{\rm E}^2\gamma_{\rm E}^3
  = \gamma^0,
\end{equation}
which satisfies 
\begin{equation}
  (\gamma_{\rm E}^5)^2 = {\bf 1}, \quad
 \{\gamma_{\rm E}^5,\gamma_{\rm E}^{\mu}\} = 0 \quad (\mu=0,\cdots, 3). 
\end{equation}

It follows that the matrices 
\begin{equation}
 S_{\rm E}^{\mu\nu} \equiv \frac{i}{4}[\gamma_{\rm E}^{\mu},\gamma_{\rm E}^{\nu}]
\end{equation}
satisfy the algebra of $SO(4)$ rotation generators 
\begin{equation}
 [S_{\rm E}^{\mu\nu},S_{\rm E}^{\rho\sigma}] = 
  i(\delta^{\nu\rho}S_{\rm E}^{\mu\sigma}-\delta^{\mu\rho}S_{\rm E}^{\nu\sigma}
  -\delta^{\nu\sigma}S_{\rm E}^{\mu\rho}+\delta^{\mu\sigma}S_{\rm E}^{\nu\rho}).
\end{equation}
Hence, the $SO(4)$ rotation for the Dirac field $\psi$ is 
\begin{equation}
 \psi \to \Lambda_{{\rm E},\frac12}\psi, \quad
  \Lambda_{{\rm E},\frac12} = 
  \exp\left[-\frac{i}{2}\omega^{\rm E}_{\mu\nu}S_{\rm E}^{\mu\nu}\right],
\end{equation}
where $\omega^{\rm E}_{\mu\nu}$ are real numbers. Since all 
$S_{\rm E}^{\mu\nu}$  are Hermitian, $\Lambda_{{\rm E},\frac12}$ is
unitary. In particular, this implies that 
\begin{eqnarray}
 \bar{\psi} \to  \bar{\psi}\Lambda_{{\rm E},\frac12}^{-1}, 
  \qquad
 \psi^{\dagger}  \to  \psi^{\dagger}\Lambda_{{\rm E},\frac12}^{-1},
\end{eqnarray}
and that both $\bar{\psi}\psi$ and
$\bar{\psi}\gamma_{\rm E}^5\psi$ ($=\psi^{\dagger}\psi$) are scalars under a $SO(4)$ transformation (see e.g. Refs.~\cite{Wetterich:2010ni,mehta}). Note also that $\bar{\psi}=\psi^{\dagger}\gamma^0$ can be written as
\begin{equation}
 \bar{\psi} = \psi^{\dagger}\gamma_{\rm E}^5.
\end{equation}

\subsubsection{Euclidean action and emergence of the Lorentzian Dirac action}\label{subsubsec2.4.3}

As in the previous sections, we will need to couple the spinor field $\psi$ to the clock field $\phi$ in order for the spinor to have an apparent Lorentzian dynamics. Starting from the Euclidean Dirac action in flat space with the metric $\delta_{\mu\nu}$, 
\begin{equation}
 \int dx^4\bar{\psi}
  \left(\frac{i}{2}\gamma_{\rm E}^{\mu}
   {\overleftrightarrow{\partial_\mu}}-m\right)\psi, \nonumber
\end{equation}
and assuming that  the clock field $\phi$ has derivative couplings to the Euclidean Dirac field $\psi$ of the form
\begin{equation}
 \int dx^4
  \delta^{\mu\nu}
  (i\bar{\psi}\gamma_{\rm E}^5{\overleftrightarrow{\partial_\mu}}\psi)\partial_{\nu}\phi, 
  \quad
 \int dx^4
  \delta^{\mu\nu}
  (i\bar{\psi}\gamma_{\rm E}^{\rho}{\overleftrightarrow{\partial_\mu}}\psi)\partial_{\rho}\phi\partial_{\nu}\phi,  \nonumber
\end{equation}
we can consider an Euclidean action for the Dirac spinor of the form
\begin{eqnarray}\label{e.spin1}
 S_{\psi} &=& \int dx^4
  \left\lbrace \bar{\psi}\left(
  \frac{i}{2}\gamma_{\rm E}^{\mu}
   {\overleftrightarrow{\partial_\mu}}-m\right)\psi\right.\\
  &+&\left. \frac{1}{2M^2}  \delta^{\mu\nu}\left[
  (i\bar{\psi}\gamma_{\rm E}^5{\overleftrightarrow{\partial_\mu}}\psi)
  - \frac{1}{M^2}
  (i\bar{\psi}\gamma_{\rm E}^{\rho}{\overleftrightarrow{\partial_\mu}}\psi)\partial_{\rho}\phi\right]\partial_{\nu}\phi
  \right\rbrace.\nonumber
\end{eqnarray}

As in the previous sections, the action $S_{\psi}$ reduces to 
\begin{equation}
 S_{\psi} = \int dx^4
  \bar{\psi} \left[ 
     \frac{i}{2}\gamma^0
   {\overleftrightarrow{\partial_0}}
   + 
   \frac{i}{2}\gamma^i
   {\overleftrightarrow{\partial_i}}
   -m\right]\psi. 
\end{equation}
The coupling to the clock field implies that $\psi$ effectively
propagates in an effective Lorentzian metric and we recover the standard
Minkowskian Dirac action~(\ref{e.diracstd}) with the usual
algebra~(\ref{e.Dmat}) for the $\gamma$-matrices.

\subsection{Massive point particle}

The dynamics of massive object is usually derived from an action defined from the length of their worldline. In order to recover a proper dynamics, we start from the Euclidean action for a point particle
$$
\frac{1}{2}\int
\left({\cal N}^{-1}\delta_{\mu\nu}
\frac{dx^{\mu}}{d\tau}\frac{dx^{\nu}}{d\tau}
- {\cal N}m^2\right) d\tau
$$
to which we add the coupling to the clock field of the form
$$
\int
{\cal N}^{-1}\partial_{\mu}\phi\partial_{\nu}\phi
\frac{dx^{\mu}}{d\tau}\frac{dx^{\nu}}{d\tau}d\tau
$$
The Euclidean action for a point particle is thus given by
\begin{eqnarray}
 S_{\rm pp} & = & 
  \frac{1}{2}\int
  \bigg[{\cal N}^{-1}
 \left(\delta_{\mu\nu}-\frac{2}{M^4}
  \partial_{\mu}\phi\partial_{\nu}\phi\right)
 \frac{dx^{\mu}}{d\tau}\frac{dx^{\nu}}{d\tau}
 \nonumber\\
 & &  - {\cal N}m^2\bigg] d\tau. 
\end{eqnarray}
The equation of motion is thus simply given by the geodesic equation for
the effective metric 
$g_{\mu\nu}^{(m)}= g_{\mu\nu}^{\rm E}-\frac{2}{M^4}\partial_\mu\phi\partial_\nu\phi$. 
It is obvious that in ${\cal M}_0$ this effective metric reduces to the
Minkowski metric $\eta_{\mu\nu}$.

\subsection{Discussion}\label{subsec2.5}

This section has provided the general construction of a mechanism that
allows for scalar, vector, and spinor fields to actually propagate in an
effective Lorentzian metric even though the underlying theory is purely
Euclidean and written in terms of the Euclidean metric
$\delta_{\mu\nu}$.  This general construction assumes the existence of a
scalar field $\phi$, called {\em clock field}, that couples to all
fields (scalar, vector and spinor fields). In particular, this implies
that we can construct the whole standard model of particle physics. 

Let us now discuss some properties and limitations of such a
construction. 
\begin{enumerate}
 \item It requires that the clock field satisfies
       $\partial_\mu\phi=\hbox{const.}\not=0$ in a region ${\cal M}_0$
       of the Euclidean space. It follows that the effective Lorentzian
       description is local and holds in ${\cal M}_0$. The properties of 
       this model when $\partial_\mu\phi$ is not constant will be
       discussed in \S~\ref{subsec5.3} below.  As we shall see in the
       next section the clock field should enjoy a shift symmetry in
       order for the system to exhibit the time translation symmetry
       after the emergence of time. In ${\cal M}_0$, both the shift
       symmetry and the translational symmetry along the direction of
       $\partial_{\mu}\phi$ are spontaneously broken, but a combination
       of them remains unbroken and is responsible for the existence of
       a conserved quantity that reduces in ${\cal M}_0$ to the usual
       notion of energy. 
 \item It is limited to classical field theory in flat space. The
       extension to curved space is discussed in \S~\ref{sec3} and
       \S~\ref{sec4} below and the quantum aspects are left for future 
       investigations. 
 \item The origin of the effective Lorentzian dynamics in ${\cal M}_0$
       can be intuitively understood for scalars and vectors. For scalars,
       the action~(\ref{e.chi1}) is equivalent to the coupling to the
       effective metric 
 \begin{equation}
   \hat g^{\mu\nu}=\delta^{\mu\nu}-\frac{2}{M^4}\delta^{\mu\alpha}\delta^{\nu\beta}\partial_\alpha\phi\partial_\beta\phi.
 \end{equation}  
       For vectors, one could have simply used a coupling to  
       $\hat g^{\mu\nu}$ and a Lagrangian of the form 
       $ \hat g^{\mu\alpha} \hat g^{\nu\beta}F_{\mu\nu}F_{\alpha\beta}$
       since the extra term quartic in $\partial_{\mu}\phi$ compared to
       the action~(\ref{e.A1}) is of the form
       $4M^{-8}\delta^{\mu\lambda}\delta^{\alpha\lambda'}\delta^{\nu\sigma}\delta^{\beta\sigma'}\partial_{\lambda}\phi\partial_{\lambda'}\phi\partial_{\sigma}\phi\partial_{\sigma'}\phi F_{\mu\nu}F_{\alpha\beta}$
       and does not contribute (note that it reduces to $4F_{00}^2=0$ in 
       ${\cal M}_0$). Hence, the apparent Lorentzian dynamics for scalars
       and vectors boils down to the fact that 
       $\hat g^{\mu\nu}\vert_{{\cal M}_0}=\eta^{\mu\nu}$. Massive point 
       particles also propagate in this metric.  
\item This interpretation cannot be extended to spinors mostly because
      of the $\gamma$-matrices, at least straightforwardly. 
\item It is however important to realize that despite this, when restricted to ${\cal M}_0$ all fields propagate in the same effective Minkowski metric so that the equivalence principle is safe in first approximation.
 \item The couplings to the clock field have been tuned in order to recover the exact Minkowski actions. For instance, the action~(\ref{e.chi1}) for a scalar field could have been chosen as
 \begin{eqnarray}\label{e.chi2}
 S_\chi &=&  \int \dd^4x \left[ - \frac{\kappa_\chi}{2}\delta^{\mu\nu}\partial_\mu\chi\partial_\nu\chi   -V(\chi)\right.\nonumber\\
 &&\left.\qquad\qquad+\frac{\alpha_\chi}{2M^4}\left( \delta^{\mu\nu}\partial_\mu\phi\partial_\nu\chi\right)^2
 \right].
\end{eqnarray}
In such a case, a Lorentzian signature is recovered only if $\alpha_\chi>\kappa_\chi>0$. In the case where these constants are not tuned, different fields can have different lightcones. This will be discussed in Section~\ref{sec5}.
 \item In the bosonic sector, since the theory is invariant
       under the Euclidean parity ($x^{\mu}\rightarrow -x^{\mu}$) as
       well as the field parity ($\phi\rightarrow -\phi$), both $P$ and
       $T$ invariances in the emergent Lorentzian theory are
       ensured. Without the field parity invariance, the $T$ invariance would be
       spontaneously broken by a non-vanishing vacuum expectation value
       (vev) of the derivative of the clock field. This explains the reason
       why we have included only quadratic terms in $\partial_\mu\phi$
       in the actions for scalars and vectors. 
\item In the fermionic sector, let us first remark that one could have constructed  $16$ independent Euclidean $\gamma$-matrices, explicitly given by
\begin{equation}
 {\bf 1}, \quad \gamma_{\rm E}^5, \quad \gamma_{\rm E}^{\mu}, \quad
\gamma_{\rm E}^5\gamma_{\rm E}^{\mu}, \quad S_{\rm E}^{\mu\nu}. 
\end{equation}
 From the Dirac spinor $\psi$, we can thus construct bilinear combinations that transform as scalars under $SO(4)$ rotations. Among them, Hermitian bilinears that do not include more than one derivative acting on spinors are the following ten possibilities
\begin{eqnarray}\label{eqn:10bilinears}
\quad\qquad\begin{array}{ccc}
 \bar{\psi}\psi,&\, & \bar{\psi}\gamma_{\rm E}^5\psi,\\
 i\bar{\psi}\gamma_{\rm E}^{\mu}{\overleftrightarrow{\partial_\mu}}\psi ,  &\,  &  \bar{\psi}\gamma_{\rm E}^5\gamma_{\rm E}^{\mu}{\overleftrightarrow{\partial_\mu}}\psi, \\
 &&\\
 (\bar{\psi}\gamma_{\rm E}^{\mu}\psi)\partial_{\mu}\phi, &\,  & (i\bar{\psi}\gamma_{\rm E}^5\gamma_{\rm E}^{\mu}\psi)\partial_{\mu}\phi, \\
\delta^{\mu\nu} (i\bar{\psi}{\overleftrightarrow{\partial_\mu}}\psi)\partial_{\nu}\phi, &\,  &  \delta^{\mu\nu}(i\bar{\psi}\gamma_{\rm E}^5{\overleftrightarrow{\partial_\mu}}\psi)\partial_{\nu}\phi, \\
 &&\\
 \delta^{\mu\nu}  (i\bar{\psi}\gamma_{\rm E}^{\rho}{\overleftrightarrow{\partial_\mu}}\psi)\partial_{\rho}\phi\partial_{\nu}\phi,&\,  & \delta^{\mu\nu}(\bar{\psi}\gamma_{\rm E}^5\gamma_{\rm E}^{\rho}{\overleftrightarrow{\partial_\mu}}\psi)\partial_{\rho}\phi\partial_{\nu}\phi.
\end{array}\nonumber
\end{eqnarray}
      The first two of the left column correspond to the standard mass
      and kinetic terms while six among the eight others describe
      possible couplings to the clock field. Among these six couplings,
      we have only used the two which were sufficent as an existence
      proof of our mechanism for Dirac spinors, namely 
      $\delta^{\mu\nu} (i\bar{\psi}\gamma_{\rm E}^5\overleftrightarrow{\partial_{\mu}}\psi)\partial_{\nu}\phi$ 
      and 
      $\delta^{\mu\nu} (i\bar{\psi}\gamma_{\rm E}^{\rho}\overleftrightarrow{\partial_{\mu}}\psi)\partial_{\rho}\phi\partial_{\nu}\phi$.
      It has to be remarked that  the second term is not CPT invariant
      after the clock field has a vev. Hence, unless the coefficient of
      this term is exactly the value shown in (\ref{e.spin1}), the CPT
      invariance is violated. We also need to emphasize that we have
      been able to construct Dirac spinor but that we also need to
      construct Majorana and Weyl spinors. This is an open problem at
      the moment. 
 \item The mass scale $M$ is related to $\partial_\mu\phi$ and is
       arbitrary. It is important to realize that it does not appear in
       the final expressions of the effective Lorentzian actions. 
 \item $X_{\rm E}$ may not be constant if
       $g_{\mu\nu}^{\rm E}\not=\delta_{\mu\nu}$ (curved space) and/or if
       $\partial_\mu\phi$ is not strictly constant in ${\cal M}_0$. This
       will be discussed in \S~\ref{sec3} and \S~\ref{sec5}. 
 \item The configuration of the clock field is not arbitrary but should
       be determined by solving the equation of motion. Since the action
       for the clock field enjoys a shift symmetry, its equation of
       motion takes the form of a current conservation. This will be
       addressed in \S~\ref{sec3}, where we will show that
       $\partial_\mu\phi=\hbox{const.}\not=0$ can be a solution,
       e.g. with $g^{\rm E}_{\mu\nu}=\delta_{\mu\nu}$. 
\end{enumerate}

\section{Gravitation and curved space}\label{sec3}

So far, our description has been restricted to the classical dynamics of
standard fields in flat spacetime. The first natural generalisation we
must consider is the way to include gravity, i.e. a theory that will
mimic or be close to general relativity. 

For this purpose, we now consider a general $4$-dimensional
Riemannian\footnote{We use the term {\it Riemannian} for a curved
spacetime with a positive definite metric and {\it Lorentzian} for a
curved spacetime with a Lorentz signature. We keep the terms 
{\it Euclidean} and {\it Minkowskian} for the analog in flat
space. However, for simplicity, we use the same subscript E for the
Riemannian and Euclidean cases.} manifold ${\cal M}$ with a positive
definite metric $g^{\rm E}_{\mu\nu}$. Again, the theory we shall
consider on this manifold does not have a microscopic concept of
time. As previously, we introduce a clock field $\phi$ and assume it
enjoys a shift symmetry that, as we have already seen, is necessary for
the system to exhibit the time translation symmetry after the emergence
of time.

\subsection{Generic couplings to the clock field}\label{subsec3.1}

In order to minimize the number of physical degrees of freedom, we
demand that the equation of motion for $\phi$ is a second-order
differential equation. Hence, the action for $\phi$ is restricted to the
Riemannian version of the Horndeski theory~\cite{Horndeski:1974wa} with
shift symmetry. Equivalently, it is given by the shift-symmetric
generalized Galileon~\cite{Deffayet:2011gz} as 
\begin{equation}
 S_g = \int \dd x^4 \sqrt{g_{\rm E}} \left(L_2 + L_3 + L_4 + L_5 \right),
\end{equation}
where the Lagrangians are explicitly given by
\begin{eqnarray}
 L_2 & = & {\cal K}(X_{\rm E}), \nonumber\\
 L_3 & = & -G_3(X_{\rm E})\nabla_{\rm E}^2\phi, \nonumber\\
 L_4 & = & G_4(X_{\rm E})R_{\rm E} - 2G_4'(X_{\rm E})
  \left[(\nabla_{\rm E}^2\phi)^2-
   (\nabla^{\rm E}_{\mu}\nabla^{\rm E}_{\nu}\phi)^2\right], \nonumber\\
 L_5 & = & -g_5 G_{\rm E}^{\mu\nu}\partial_{\mu}\phi\partial_{\nu}\phi
  + \tilde{G}_5(X_{\rm E})G_{\rm E}^{\mu\nu}\nabla^{\rm E}_{\mu}\nabla^{\rm E}_{\nu}\phi
  \nonumber\\
 & & 
  + \frac{1}{3}\tilde{G}'_5(X_{\rm E})
  \left[(\nabla_{\rm E}^2\phi)^3
   -3(\nabla_{\rm E}^2\phi)(\nabla^{\rm E}_{\mu}\nabla^{\rm E}_{\nu}\phi)^2\right.\nonumber\\
  && \left.+ 2(\nabla^{\rm E}_{\mu}\nabla^{\rm E}_{\nu}\phi)^3\right].
\end{eqnarray}
Here, $\nabla^{\rm E}_{\mu}$, $R_{\rm E}$ and $G_{\rm E}^{\mu\nu}$ are the covariant
derivative associated with the Riemannian metric $g^{\rm E}_{\mu\nu}$, its
Ricci scalar and Einstein tensor. The coefficient $g_5$ is a constant and
${\cal K}(X_{\rm E})$, $G_{3,4}(X_{\rm E})$ and 
$\tilde{G}_5(X_{\rm E})$ are arbitrary functions of $X_{\rm E}$ 
and a prime refers to a derivative with respect to $X_{\rm E}$ that is
defined as 
\begin{eqnarray}\label{e.33}
 X_{\rm E} & \equiv &   g_{\rm E}^{\mu\nu}\partial_{\mu}\phi\partial_{\nu}\phi.
\end{eqnarray}
We use the following short-hand notations
\begin{eqnarray}
  \nabla_{\rm E}^2\phi & \equiv &
  g_{\rm E}^{\mu\nu}\nabla^{\rm E}_{\mu}\nabla^{\rm E}_{\nu}\phi, \\
 (\nabla^{\rm E}_{\mu}\nabla^{\rm E}_{\nu}\phi)^2 & \equiv & 
  g_{\rm E}^{\nu\rho}g_{\rm E}^{\sigma\mu}
  (\nabla^{\rm E}_{\mu}\nabla^{\rm E}_{\nu}\phi)
  (\nabla^{\rm E}_{\rho}\nabla^{\rm E}_{\sigma}\phi), \nonumber\\
 (\nabla^{\rm E}_{\mu}\nabla^{\rm E}_{\nu}\phi)^3 & \equiv &
  g_{\rm E}^{\nu\rho}g_{\rm E}^{\sigma\alpha}g_{\rm E}^{\beta\mu}
  (\nabla^{\rm E}_{\mu}\nabla^{\rm E}_{\nu}\phi)
  (\nabla^{\rm E}_{\rho}\nabla^{\rm E}_{\sigma}\phi)
  (\nabla^{\rm E}_{\alpha}\nabla^{\rm E}_{\beta}\phi),\nonumber
\end{eqnarray}
where $g_{\rm E}^{\mu\nu}$ is the inverse of $g^{\rm E}_{\mu\nu}$.

For the effective equations, i.e. once the Lorentzian structure and the
notion of time have emerged, we would like to ensure that the system is
invariant not only under  time translation but also under
CPT~\footnote{As we have seen in the previous section, this requirement 
is not obviously fulfilled for spinors without fine-tuning. An
additional mechanism is needed to naturally ensure the CPT invariance
for spinors. In the present article we shall thus focus on the bosonic
sector.}. 

For this reason, we require that besides the shift symmetry
($\phi\to \phi + {\rm const.}$) the theory also enjoys a $Z_2$ symmetry
($\phi\to -\phi$) for the clock field action. With these symmetries, the
action reduces to   
\begin{eqnarray}   \label{eqn:action-gravity-pre} 
 S_g & = & \int \dd x^4 \sqrt{g_{\rm E}}
  \bigg\{ G_4(X_{\rm E})R_{\rm E}  - g_5G_{\rm E}^{\mu\nu}\partial_{\mu}\phi\partial_{\nu}\phi 
   + {\cal K}(X_{\rm E})\nonumber\\
  &&\quad- 2G_4'(X_{\rm E})
   \left[(\nabla_{\rm E}^2\phi)^2-
   (\nabla^{\rm E}_{\mu}\nabla^{\rm E}_{\nu}\phi)^2\right]\bigg\},
\end{eqnarray}
since only $L_2$, $L_4$ and the first  term of $L_5$ can contribute.

It is easy to show that the constant $g_5$ in the action
(\ref{eqn:action-gravity-pre}) can be absorbed into the redefinition of 
$G_4(X_E)$ up to a boundary term. Hence, by setting $g_5=0$, hereafter 
we consider the Riemannian gravity action of the form 
\begin{eqnarray}   \label{eqn:action-gravity}
 S_g & = & \int \dd x^4 \sqrt{g_{\rm E}}
  \bigg\{ G_4(X_{\rm E})R_{\rm E} 
   + {\cal K}(X_{\rm E})\nonumber\\
  &&\quad- 2G_4'(X_{\rm E})
   \left[(\nabla_{\rm E}^2\phi)^2-
   (\nabla^{\rm E}_{\mu}\nabla^{\rm E}_{\nu}\phi)^2\right]\bigg\}. 
\end{eqnarray}

\subsection{Action for the gravitational sector}\label{subsec3.2}

Following the logic developed in Section~\ref{sec2}, we restrict our analysis to a region ${\cal M}_0$ in which $X_{\rm E}>0$ so that we can define a preferred direction, that we shall call $t$, defined as in Eq.~(\ref{e.time0}), 
\begin{equation}\label{e.time}
 t\equiv \frac{\phi}{M^2},
\end{equation}
that is chosen as one of coordinates of the 4-dimensional Riemannian
manifold. We refer to such a coordinate choice (\ref{e.time})  as {\it
unitary gauge}. 

\subsubsection{Decomposition of the Riemannian metric}

One can then introduce a set of three other independent coordinates $x^i$ ($i=1,2,3$) so that the Riemannian metric is decomposed as
\begin{equation}
 g^{\rm E}_{\mu\nu}\dd x^{\mu}\dd x^{\nu} = 
  N_{\rm E}^2 \dd t^2 
  + \gamma_{ij}(\dd x^i+N^i\dd t)(\dd x^j+N^j\dd t), 
  \label{eqn:EuclideanADM}
\end{equation}
where the lapse $N_{\rm E}$ is given, thanks to Eq.~(\ref{e.time}), by
\begin{eqnarray}\label{def:NE}
 N_{\rm E} & \equiv & \frac{1}{\sqrt{g_{\rm E}^{tt}}}
  = \frac{M^2}{\sqrt{X_{\rm E}}}.
\end{eqnarray}
The 3-metric $\gamma_{ij}$ is given by
\begin{eqnarray}
 \gamma_{ij}  \equiv g^{\rm E}_{ij},
\end{eqnarray} 
and $\gamma^{ij}$ is its inverse. To finish,  the shift vector $N^i$ is
given by 
\begin{eqnarray}
 N^i & \equiv  \gamma^{ij}g^{\rm E}_{tj}.
\end{eqnarray}  
One can then easily check that the inverse Riemannian metric is given by 
\begin{eqnarray}
 g_{\rm E}^{tt} & = & \frac{1}{N_{\rm E}^2}, \nonumber\\
 g_{\rm E}^{t i} & = & g_{\rm E}^{it} = -\frac{N^i}{N_{\rm E}^2},
  \nonumber\\
 g_{\rm E}^{ij} & = & \gamma^{ij} + \frac{N^iN^j}{N_{\rm E}^2}. 
\end{eqnarray}

\subsubsection{Riemannian geometrical quantities}

With the decomposition (\ref{eqn:EuclideanADM}), it is straightforward to show that the Einstein-Hilbert term reduces to
\begin{eqnarray}
 \sqrt{g_{\rm E}}R_{\rm E} & = & 
  N_{\rm E}\sqrt{\gamma}(-K_{\rm E}^{ij}K^{\rm E}_{ij}+K_{\rm E}^2+R^{(3)})
  \nonumber\\
 & & 
  - 2\partial_i(\sqrt{\gamma}\gamma^{ij}\partial_jN_{\rm E})
  - 2\partial_t(\sqrt{\gamma}K_{\rm E}) 
  \nonumber\\
 & & 
  + 2\partial_i(\sqrt{\gamma}N^iK_{\rm E}),
\end{eqnarray}
in terms of the extrinsic curvature of the constant-$t$ hypersurface, $K^{\rm E}_{ij}$, defined by 
\begin{equation}
 K^{\rm E}_{ij} \equiv \frac{1}{2N_{\rm E}}
  (\partial_{t}\gamma_{ij}-D_iN_j-D_jN_i),
\end{equation}
where $D_i$ is the spatial covariant derivative compatible with
$\gamma_{ij}$, and $R^{(3)}$ is its Ricci scalar. We have used the
notations 
$K_{\rm E}^{ij}\equiv \gamma^{ik}\gamma^{jl}K^{\rm E}_{kl}$,
$K_{\rm E}\equiv\gamma^{ij}K^{\rm E}_{ij}$, 
and $N_i\equiv\gamma_{ij}N^j$.

\subsubsection{Riemannian action in ${\cal M}_0$}

With the use of the quantities introduced above, the Riemannian action~(\ref{eqn:action-gravity}) takes the form
\begin{eqnarray}
 S_g & = & \int \dd t \dd x^3 N_{\rm E} \sqrt{\gamma}
 \left\{
  - G_4(K_{\rm E}^{ij}K^{\rm E}_{ij}-K_{\rm E}^2) \right. \nonumber\\
 & & \left.
  \qquad\qquad
  +  G_4 R^{(3)} + L_{\phi}\right\},
\end{eqnarray} 
where the Lagrangian $ L_{\phi}$ is given by
\begin{eqnarray}\label{e.316}
  L_{\phi} & = & -2(\partial^{\rm E}_{\perp}\partial^{\rm E}_{\perp}+D^2)G_4
  - 2G_4'\left[(\nabla_{\rm E}^2\phi)^2 \right. \nonumber\\
 & & \left.
	 -(\nabla_{\rm E}^{\mu}\nabla_{\rm E}^{\nu}\phi)
	 (\nabla^{\rm E}_{\mu}\nabla^{\rm E}_{\nu}\phi)
       \right] + {\cal K}(X_{\rm E}), 
\end{eqnarray}
in which the 3-dimensional Laplacian is defined as usual as 
$D^2\equiv\gamma^{ij}D_iD_j$. The perpendicular derivative 
$\partial^{\rm E}_{\perp}$ is defined in terms of the unit vector normal
to the constant $\phi$ hypersurfaces,
$n^{\rm E}_\mu=\partial_\mu\phi/\sqrt{X_E}$, as 
\begin{equation}\label{e.317}
 \partial^{\rm E}_{\perp} \equiv 
 n_{\rm E}^{\mu}\partial_{\mu} \equiv
  \frac{1}{N_{\rm E}}\left(\partial_t-N^i\partial_i\right),
\end{equation}
with $n_{\rm E}^\mu=g_{\rm E}^{\mu\nu}n^{\rm E}_\mu$.

In order to further simplify $L_{\phi}$, note that 
$\nabla^{\rm E}_{\mu}\nabla^{\rm E}_{\nu}\phi = -M^2\Gamma^t_{E\mu\nu}$
in terms of the Christoffel symbols for the metric $g^{\rm E}_{\mu\nu}$, 
$\Gamma^{\rho}_{{\rm E}\mu\nu}$.  Its components are explicitly given
by 
\begin{eqnarray}
 \phi^{\rm E}_{;ij} & \equiv & 
  \nabla^{\rm E}_i\nabla^{\rm E}_j\phi = 
  \sqrt{X_{\rm E}}K^{\rm E}_{ij}, \nonumber\\
 \phi^{\rm E}_{;\perp i} & \equiv & \phi^{\rm E}_{;i\perp} \equiv
  n_{\rm E}^{\mu}\nabla^{\rm E}_{\mu}\nabla^{\rm E}_i\phi = 
  \frac{1}{2}\sqrt{X_{\rm E}}\partial_i\ln X_{\rm E}, \nonumber\\
 \phi^{\rm E}_{;\perp\perp} & \equiv &
  n_{\rm E}^{\mu}n_{\rm E}^{\nu}\nabla^{\rm E}_{\mu}\nabla^{\rm E}_{\nu}\phi = 
   \frac{1}{2}\sqrt{X_{\rm E}}\partial^{\rm E}_{\perp}\ln X_{\rm E}. 
\end{eqnarray}
It implies that the term $(\nabla_{\rm E}^2\phi)^2 -(\nabla_{\rm E}^{\mu}\nabla_{\rm E}^{\nu}\phi) (\nabla^{\rm E}_{\mu}\nabla^{\rm E}_{\nu}\phi)$ appearing in Eq.~(\ref{e.316}) takes the form
$$
(\gamma^{ij}\gamma^{kl}-\gamma^{ik}\gamma^{jl})
 \phi^{\rm E}_{;ij}\phi^{\rm E}_{;kl}
 + 2\gamma^{ij}
 \left(
  \phi^{\rm E}_{;\perp\perp}\phi^{\rm E}_{;ij}
  - \phi^{\rm E}_{;\perp i}\phi^{\rm E}_{;\perp j}\right)
$$
and thus reduces to
$$
-X_{\rm E}(K_{\rm E}^{ij}K^{\rm E}_{ij}-K_{\rm E}^2)+K_{\rm E}\partial^{\rm E}_{\perp}X_{\rm E} -\frac{1}{2}X_{\rm E}(D_i\ln X_{\rm E})^2.
$$
Inserting this into Eq.~(\ref{e.316}), it follows that $L_\phi$ takes
the form 
\begin{equation}
 L_{\phi} = 2G_4'X_{\rm E}(K_{\rm E}^{ij}K^{\rm E}_{ij}-K_{\rm E}^2) 
  + {\cal K}(X_{\rm E})
 + \frac{\Delta_{\phi}}{N_{\rm E}\sqrt{\gamma}},
\end{equation}
where the last term is given by
\begin{eqnarray}
 \Delta_{\phi} & = & 
  -2\partial_t\left(\sqrt{\gamma}\partial^{\rm E}_{\perp}G_4\right)
  + 2\partial_i\left(\sqrt{\gamma}N^i\partial^{\rm E}_{\perp}G_4\right)
  \nonumber\\
 & & 
  - 2\partial_i\left(\sqrt{\gamma}\gamma^{ij}N_{\rm E}\partial_jG_4\right)
\end{eqnarray}
and is a total derivative. We finally obtain the expression of the
Riemannian action 
\begin{eqnarray}
 S_g & = & \int \dd t \dd x^3 N_{\rm E} \sqrt{\gamma}
 \bigg\{
  (2G_4'X_{\rm E} - G_4)
  (K_{\rm E}^{ij}K^{\rm E}_{ij}-K_{\rm E}^2)\nonumber\\
 & & 
  +  G_4 R^{(3)} + {\cal K}(X_{\rm E})\bigg\}, 
 \label{eqn:action-gravity2}
\end{eqnarray} 
where it is understood that $X_{\rm E}$ defined in Eq.~(\ref{e.33}) is given by
$$
 X_{\rm E}=\frac{M^4}{N_{\rm E}^2}
$$ 
and that the time coordinate is fixed according to the unitary gauge (\ref{e.time}). 

\subsubsection{Lorentzian metric}

We now introduce a Lorentzian metric $g_{\mu\nu}$ and decompose it as
\begin{equation}  \label{eqn:LorentzianMetric}
 g_{\mu\nu}\dd x^{\mu}\dd x^{\nu}  = -N^2\dd t^2 +\gamma_{ij}(\dd x^i+N^i\dd t)(dx^j+N^j\dd t),
\end{equation}
where the lapse $N$ is defined by
\begin{equation}
 N \dd N=-N_{\rm E}\,\dd N_{\rm E}
\end{equation}
so that the Riemannian and Lorentzian lapses are related to each other
as 
\begin{equation}\label{e.234}
 N = \sqrt{N_c^2-N_{\rm E}^2},
\end{equation} 
$N_c$ being an arbitrary positive constant. As above, we can define the extrinsic curvature of this metric as
\begin{eqnarray}
 K_{ij} & \equiv &  \frac{1}{2N}  (\partial_{t}\gamma_{ij}-D_iN_j-D_jN_i)
\end{eqnarray} 
 and $ K^{ij} \equiv \gamma^{ik}\gamma^{jl}K_{kl}$, $K \equiv\gamma^{ij}K_{ij}$. It is related to the Riemannian extrinsic curvature by
\begin{eqnarray} 
   K_{ij}  &=& \frac{N_{\rm E}}{N}K^{\rm E}_{ij}, \nonumber\\
 K^{ij} & = & \frac{N_{\rm E}}{N}K_{\rm E}^{ij}, \nonumber\\
 K &= &  \frac{N_{\rm E}}{N}K_{\rm E}.
\end{eqnarray}
The Ricci scalar of $g_{\mu\nu}$ can be expressed in terms of the extrinsic curvature by the well-known formula
\begin{equation}\label{e.327}
\sqrt{-g}R =  N \sqrt{\gamma}
  \left[K^{ij}K_{ij}-K^2 + R^{(3)}\right] - \Delta
 \end{equation}
with $\Delta=2\partial_i\left(\sqrt{\gamma}\gamma^{ij}\partial_jN\right)-2\partial_t\left(\sqrt{\gamma}K\right)+ 2\partial_i\left(\sqrt{\gamma}N^iK\right)$  and $g=\det g_{\mu\nu}$.

\subsubsection{Lorentzian action in unitary gauge}

In the unitary gauge we have been using so far, the action~(\ref{eqn:action-gravity2}) is now rewritten as 
\begin{eqnarray}
 S_g  & = & \int \dd t \dd x^3 N \sqrt{\gamma}
  \left\{\left[f(X)-2Xf'(X)\right](K^{ij}K_{ij}-K^2)
	       \right. \nonumber\\
 & & \left.
  \qquad\qquad\qquad\qquad
  + f(X) R^{(3)} +  P(X)\right\},
 \label{eqn:action-gravity3}
\end{eqnarray}
where the functions $f$ and $P$ are defined by
\begin{eqnarray}  \label{eqn:def-X-f-P}
  f(X) & \equiv & 
  \frac{N_{\rm E}}{N}G_4(X_E),
  \nonumber\\
 f'(X) & \equiv & \frac{\dd f(X)}{\dd X}, \nonumber\\
 P(X) & \equiv & \frac{N_{\rm E}}{N}{\cal K}(X_{\rm E})
\end{eqnarray}
in terms of
\begin{eqnarray}
 X & \equiv & \frac{M^4}{N^2}. 
\end{eqnarray}
To show the equivalence between Eq.~(\ref{eqn:action-gravity2}) and Eq.~(\ref{eqn:action-gravity3}), we have noted that Eq.~(\ref{e.234}) implies that
\begin{equation}
 \frac{1}{X} + \frac{1}{X_{\rm E}} = \frac{N_c^2}{M^4}, \quad
 \frac{\dd X}{\dd X_{\rm E}} = -\frac{X^2}{X_{\rm E}^2}.
\end{equation}

Now, using the property~(\ref{e.327}), the action~(\ref{eqn:action-gravity3}) can be further simplified to
\begin{eqnarray}
 S_g & = & \int \dd t \dd x^3 N \sqrt{\gamma}
 \left\{
  f(X) R - 2Xf'(X)(K^{ij}K_{ij}-K^2)
				  \right. \nonumber\\
 & & \left.
  + f'(X)\left[\frac{(D_iX)^2}{X} + 2K\partial_{\perp}X\right]
  + P(X)\right\}, 
 \label{eqn:action-gravity4}
\end{eqnarray} 
where the perpendicular derivative $ \partial_{\perp}$ is defined similarly as Eq.~(\ref{e.317}) in terms of the normal vector to the constant $\phi$ hypersurfaces, $n_\mu=\partial_\mu\phi/\sqrt{X}$, as
\begin{equation}\label{e.334}
 \partial_{\perp} = n^{\mu}\partial_{\mu} 
  = \frac{1}{N}(\partial_t-N^i\partial_i),
\end{equation} 
with $n^\mu=g^{\mu\nu}n_\mu$. 

\subsubsection{Covariant expression}

In the previous section the action has been derived assuming that the time coordinate was fixed according to the unitary gauge (\ref{e.time}). 

The action~(\ref{eqn:action-gravity4}) can be rewritten in a covariant way by noting that $\nabla_{\mu}\nabla_{\nu}\phi = -M^2\Gamma^t_{\mu\nu}$, where $\nabla_{\mu}$ is the covariant derivative compatible with the Lorentzian metric $g_{\mu\nu}$ and $\Gamma^{\rho}_{\mu\nu}$  are its Christoffel symbols for $g_{\mu\nu}$. Concretely, its components are given by
\begin{eqnarray}
 \phi_{;ij} & \equiv & 
  \nabla_i\nabla_j\phi = 
  -\sqrt{X}K_{ij}, \nonumber\\
 \phi_{;\perp i} & \equiv & \phi_{;i\perp} \equiv
  n^{\mu}\nabla_{\mu}\nabla_i\phi = 
  \frac{1}{2}\sqrt{X}\partial_i\ln X, \nonumber\\
 \phi_{;\perp\perp} & \equiv &
  n^{\mu}n^{\nu}\nabla_{\mu}\nabla_{\nu}\phi = 
   \frac{1}{2}\sqrt{X}\partial_{\perp}\ln X. 
\end{eqnarray}
Hence,  the term $(\nabla^2\phi)^2 -(\nabla^{\mu}\nabla^{\nu}\phi) (\nabla_{\mu}\nabla_{\nu}\phi)$ can be expressed as
$$ 
(\gamma^{ij}\gamma^{kl}-\gamma^{ik}\gamma^{jl}) \phi_{;ij}\phi_{;kl}
 - 2\gamma^{ij}
 \left(
  \phi_{;\perp\perp}\phi_{;ij}
  - \phi_{;\perp i}\phi_{;\perp j}\right),
$$  
which reduces to  
$$
-X(K^{ij}K_{ij}-K^2)+K\partial_{\perp}X
  +\frac{1}{2}\frac{(D_iX)^2}{X}.
$$
Finally, the Lorentzian action takes the form
\begin{eqnarray} \label{eqn:action-gravity5}
 S_g & = & \int \dd x^4 \sqrt{-g}
 \left\{
  f(X) R   + 2f'(X)
  \left[(\nabla^2\phi)^2
   \right. \right. \nonumber\\
 & & \left.\left.
   -(\nabla^{\mu}\nabla^{\nu}\phi)
   (\nabla_{\mu}\nabla_{\nu}\phi) \right]
  + P(X)\right\}. 
\end{eqnarray} 
Whilst this form of the action was derived assuming the unitary gauge~(\ref{e.time}), it can become manifestly covariant by promoting $X$ to a scalar defined by
\begin{equation}
 X = -g^{\mu\nu}\partial_{\mu}\phi\partial_{\nu}\phi.
\end{equation}
It is thus well-defined without the unitary gauge condition. Actually,
the covariant action (\ref{eqn:action-gravity5}) is a special case of
the covariant Galileon considered in Ref.~\cite{Deffayet:2011gz} coupled
to the Lorentzian metric $g_{\mu\nu}$. In particular, the equations of
motion are second order (see Ref.~\cite{christos} for comparison).

\subsection{Correspondence}

The derived Lorentzian theory (\ref{eqn:action-gravity5}) and the parent Riemannian theory (\ref{eqn:action-gravity}) are related to each other by the following relations. 
\begin{eqnarray}
 g_{\mu\nu} & = & g^{\rm E}_{\mu\nu}
  - \frac{\partial_{\mu}\phi\partial_{\nu}\phi}{X_c}, \nonumber\\
 g^{\mu\nu} & = & g_{\rm E}^{\mu\nu}
  + \frac{g_{\rm E}^{\mu\rho}g_{\rm E}^{\nu\sigma}
  \partial_{\rho}\phi\partial_{\sigma}\phi}{X_c-X_{\rm E}}, \nonumber\\
 \frac{1}{X} & = & \frac{1}{X_c}-\frac{1}{X_{\rm E}}, \nonumber\\
 \frac{f(X)}{\sqrt{X}} & = & 
  \frac{G_4(X_{\rm E})}{\sqrt{X_{\rm E}}}, \nonumber\\
 \frac{P(X)}{\sqrt{X}} & = & \frac{{\cal K}(X_{\rm E})}{\sqrt{X_{\rm E}}},
  \label{eqn:gravity-correspondence}
\end{eqnarray}
where $X_c$ is an arbitrary positive constant  given by
\begin{equation}
 X_c=\frac{M^4}{N_c^2}.
\end{equation}
These relations are well-defined even without the unitary gauge
condition as far as $X_{\rm E}/X_c$ is large enough. It is
straightforward to express various quantities defined in the Lorentzian
theory in terms of those in the Riemannian theory.

\subsection{Stability analysis}

We now analyze the stability of a general flat Friedmann-Lema\^{\i}tre (FL) background using the Lorentzian action (\ref{eqn:action-gravity5}) with the Lorentzian ADM decomposition (\ref{eqn:LorentzianMetric}). 

\subsubsection{Cosmological background}

We consider a flat FL background spacetime for which the metric in
cosmic time reduces to 
\begin{equation}
 N = 1, \quad N_i=0, \quad \gamma_{ij}=a(t)^2\delta_{ij},
\end{equation}
where $a$ is the scale factor, and for which the clock field
$\phi =\phi_0(t)$.

The action~(\ref{eqn:LorentzianMetric}) being invariant under a constant
shift of the clock field $\phi$, there is a conserved current associated
with the shift symmetry so that the equation of motion for $\phi$ takes
the form 
\begin{equation}\label{e.340}
\dot{J}_{\phi} + 3H J_{\phi} = 0, 
\end{equation}
where $H=\dot a/a$ is the Hubble function and 
\begin{equation}
 J_{\phi} \equiv 
  \left[ P'_0 + 6H^2(2X_0f''_0+f'_0)\right]\dot{\phi}_0. 
\end{equation}
We are using notations according to which
\begin{eqnarray}
  X_0 = \dot{\phi}_0^2, \
  P^{(n)}_0 = P^{(n)}(X_0), \
  f_0^{(n)} = f^{(n)}(X_0),
\end{eqnarray}
where $n$ stands for the order of the derivation. Thus,
Eq.~(\ref{e.340}) implies that $J_{\phi}$ decays as
$J_{\phi}\propto 1/a^3$.

By using the correspondence (\ref{eqn:gravity-correspondence}), $J_{\phi}$ can be expressed in the language of the Riemannian theory as 
\begin{eqnarray}
 J_{\phi}\dot{\phi}_0 & = & 
  \left\{
   \left[4G''_4X_{\rm E}^2+4G'_4X_{\rm E}-G_4\right]r^{3/2}
   \right.\nonumber\\
 & & \left.
   +  \left[2G'_4X_{\rm E}-G_4\right]r^{1/2}
  \right\}\times 3H^2  \nonumber\\
 & & 
  + \frac{1}{2}\left[({\cal K}-2{\cal K}'X_{\rm E})r^{1/2}+\frac{{\cal K}}{r^{1/2}}\right],
\end{eqnarray}
where a prime in the right hand side represents derivative with respect
to $X_{\rm E}$, and where the ratio $r$ is defined by
\begin{equation}  \label{def:r}
 r \equiv \frac{X_{\rm E}}{X} = \frac{X_{\rm E}}{X_c} - 1.
\end{equation}
 From Eq.~(\ref{e.234}), we have that $N_c^2>(N_{\rm E}^2,N^2)$ which implies that $r>0$.

The equation of motion for the metric reduces, as usual, to the
Friedmann equation that takes the form 
\begin{equation}
 3M_{\rm eff}^2 H^2 = 2J_{\phi}\dot{\phi}_0-P_0, 
  \label{eqn:Friedmanneq}
\end{equation}
where the effective mass scale is defined by
\begin{equation}
 M_{\rm eff}^2 \equiv 2(f_0-2X_0f'_0).
\end{equation}

To understand the qualitative behaviour of the system, let us suppose that $H^2/M^2\ll 1$ and Taylor expand $P'(X)$ and $f'(X)$ around a local minimum of $P(X)$ (which we denote as $X\equiv qM^4$)  as 
\begin{equation}
 P'(X) = p_2\delta + {\cal O}(\delta^2), \ 
  f'(X) = \frac{f_1+f_2\delta}{M^2} +{\cal  O}(\delta^2),
\end{equation}
where $q$, $p_2$ and $f_{1,2}$ are dimensionless constants of order
unity, and $\delta\equiv\frac{X}{M^4}-q$ is a small
quantity. Accordingly,  
\begin{equation}
 J_{\phi} = 
  \left[p_2\delta + 6\frac{H^2}{M^2}(2f_2q+f_1) 
   + {\cal O}\left(\frac{H^2}{M^2}\delta,\delta^2\right)\right]\dot{\phi}_0. 
\end{equation}
As already stated above, $J_{\phi}$ behaves as $\propto 1/a^3\to 0$
($a\to\infty$). Hence, apart from the trivial behavior with
$\dot{\phi}_0\to 0$, the system has a non-trivial attractor with
$\delta+{\cal O}(H^2/M^2)\propto 1/a^3\to 0$. This implies that 
$\dot{\phi}_0\to\sqrt{q}M^2[1+{\cal O}(H^2/M^2)]$ and that 
$M_{\rm eff}^2$ and $P_0$ approach constant values up to 
${\cal O}(H^2/M^2)$ corrections. Therefore, Eq.~(\ref{eqn:Friedmanneq})
is no more than the standard Friedmann equation for a universe
containing a pressureless fluid (from $J_\phi\propto 1/a^3$) and a
cosmological constant (from $P_0\to const$). This behavior is similar to
the one obtained in ghost condensate
models~\cite{ArkaniHamed:2003uy,ArkaniHamed:2005gu}. To be consistent
with the cosmic expansion as understood today, we need to have 
\begin{equation}
 P_0<0.
\end{equation}
More precisely, we even need $P_0$ to be tuned so that
\begin{equation}\label{e.cosmo1}
 P_0\sim -3\Omega_{\Lambda0} M_{\rm eff}^2H_0^2\sim -2.1 M_{\rm eff}^2H_0^2,
\end{equation}
where $\Omega_{\Lambda0}\sim0.7$ is the standard density parameter for
the cosmological constant. Note also that since $J_\phi$ contributes to
the dark matter component, it has to be bounded so that we shall require 
\begin{equation}\label{e.cosmo2}
 \frac{2}{3}\frac{J_{\phi_0}}{M^2_{\rm eff}}\sqrt{q}\frac{M^2}{H_0^2} \leq \Omega_{\rm m0} \sim 0.3,
\end{equation}
today. Note that the term can even be negative at the expense of
introducing more dark matter. These two last bounds are the only indicative
form that can be derived from cosmology, but a full cosmological
analysis will be presented elsewhere.

\subsubsection{Tensor perturbations}

We now consider tensor (T) perturbations around the FL background so that the metric is given by
\begin{eqnarray}
 & & N = 1, \ 
  N_i = 0, \
  \gamma_{ij} = a(t)^2 \left[\hbox{e}^h\right]_{ij},
\end{eqnarray} 
where $h_{ij}$ is transverse and traceless (i.e. $\partial_{i}h_{k}^i=0=\delta^{ij}h_{ij}$). We still have that $\phi = \phi_0(t)$.

In Fourier space, the quadratic action for each polarization of the tensor mode is given by 
\begin{equation}\label{e.Tmodes}
 \delta S^{(2)}_{{\rm T}, {\bm k}} = \frac{1}{8}\int \dd t a^3 
  \left[M_{\rm eff}^2\dot{h}_{\bm k}^2 - 
   2f_0\frac{{\bm k}^2}{a^2}h_{\bm k}^2 \right].  
\end{equation}
Note that this result can be easily inferred from the expression~(\ref{eqn:action-gravity3}). Hence, the stability of the tensor sector requires that 
\begin{equation}
 M_{\rm eff}^2 > 0, \qquad f_0 > 0.
\end{equation}

By using the correspondence~(\ref{eqn:gravity-correspondence}),  $M_{\rm eff}^2$ and $f_0$ are expressed in the language of the Riemannian theory as 
\begin{eqnarray}
 M_{\rm eff}^2 & = & 2(2G'_4X_{\rm E}-G_4)\sqrt{r},
  \nonumber\\ 
 f_0 & = & \frac{1}{\sqrt{r}}G_4,
\end{eqnarray}
where a prime in the right hand side of these expressions represents
derivative with respect to $X_{\rm E}$ and $r$ is defined in
Eq.~(\ref{def:r}). Thus the stability of the tensor sector gives the
constraint 
\begin{equation}\label{e.tensorC}
 2G'_4X_E > G_4 > 0.
\end{equation}

\subsubsection{Scalar perturbations}

For scalar perturbations around the FL background, the metric in the unitary gauge is given by
\begin{eqnarray}
 N = 1 + \alpha, \quad
 N_i = \partial_i \beta, \quad
  \gamma_{ij} = a(t)^2 \hbox{e}^{2\zeta}\delta_{ij}, 
\end{eqnarray} 
and, by definition, $\phi = \phi_0(t)$.

It is then straightforward to calculate the quadratic perturbed action since the time derivatives of $\alpha$ and $\beta$ do not appear in the action. Thus, the equations of motion for $\alpha$ and $\beta$ become constraint equations. After solving for those constraint equations with respect to $\alpha$ and $\beta$, one gets that the perturbed action for $\zeta$, in Fourier space, is
\begin{equation}  \label{eqn:quadratic-action-scalar}
 \delta S^{(2)}_{{\rm S}, {\bm k}} = \frac{1}{2}\int \dd t a^3 
  \left[{\cal A}\dot{\zeta}_{\bm k}^2
   - {\cal B}\frac{{\bm k}^2}{a^2}\zeta_{\bm k}^2 \right],
\end{equation}
where ${\cal A}$ and ${\cal B}$ are given by
\begin{eqnarray}
 {\cal A} & = & \frac{M_{\rm eff}^2}{H^2{\cal G}^2}
  \left(6+M_{\rm eff}^2{\cal F}\right),
  \nonumber\\
 {\cal B} & = & \frac{1}{a}
  \frac{\dd}{\dd t}\left(\frac{aM_{\rm eff}^4}{H{\cal G}^2}\right)+ 4f_0,
\end{eqnarray} 
with ${\cal F}$ and ${\cal G}$ given by
\begin{eqnarray}
 {\cal F} & = & P''_0X_0^2 + \frac{1}{2}J_{\phi}\dot{\phi}_0
  \nonumber\\
 & & 
  + 3H^2\left[4f'''_0X_0^3+14f''_0X_0^2+6f'_0X_0-f_0\right],
  \nonumber\\
 {\cal G} & = & 
  4f''_0X_0^2+4f'_0X_0 -f_0. 
\end{eqnarray}
The quadratic action (\ref{eqn:quadratic-action-scalar}) agrees with a special case of the action derived in Ref.~\cite{Kobayashi:2011nu}. The stability of scalar perturbations requires that 
\begin{equation}\label{e.scalarC2}
 {\cal A}>0, \quad {\cal B}>0.
\end{equation} 
By using the correspondence (\ref{eqn:gravity-correspondence}), this sets 2 other constraints on the Riemannian theory since ${\cal F}$ and ${\cal G}$ can be expressed in the language of the Riemannian theory as 
\begin{eqnarray}
{\cal F} & = & 
 - \bigg\{
  \left[4G'''_4X_{\rm E}^3+18G''_4X_{\rm E}^2 
   + 9G'_4X_{\rm E} - \frac{3}{2}G_4  \right]r^{5/2}\nonumber\\
  & & 
   +\left[ 2G''_4X_{\rm E}^2 + 2G'_4X_{\rm E}
     - \frac{1}{2}G_4\right]r^{3/2}\bigg\}\times 3H^2
   \nonumber\\
 & & 
  + \left({\cal K}''X_{\rm E}^2+{\cal K}'X_{\rm E}-\frac{1}{4}{\cal K}\right)r^{3/2}
 \nonumber\\
 & &  
  -\frac{1}{4}({\cal K}-2{\cal K}'X_{\rm E})r^{1/2},
  \nonumber\\
 {\cal G} & = & 
  \left[ 4G''_4X_{\rm E}^2 + 4G'_4X_{\rm E} 
	      - G_4 \right]r^{3/2},
\end{eqnarray}
where a prime in the right hand side of these expressions represents
derivative with respect to $X_{\rm E}$ and $r$ is defined in
Eq.~(\ref{def:r}). 

\subsection{Summary}

Starting from the Riemannian action~(\ref{eqn:action-gravity}) for a
positive definite metric $g^{\rm E}_{\mu\nu}$, we have been able to
derive an action for a Lorentzian metric $g_{\mu\nu}$. The key
ingredient is the coupling of the Einstein tensor of 
$g_{\mu\nu}^{\rm E}$ to the clock field. In ${\cal M}_0$, the dynamics
of $g_{\mu\nu}$ is dictated by the covariant action
(\ref{eqn:action-gravity5}), which is a special case of the covariant
Galileon considered in Ref.~\cite{Deffayet:2011gz}. (See
Ref.~\cite{Nicolis:2008in} for the original Galileon theory.)

The theory has 2 free functions, ${\cal K}$ and $G_4$, and we have shown
that the stability of the FL spacetime with respect of both scalar and
tensor perturbations at linear level sets 4 constraints on these
quantities. An extra-constraint appears from the requirement that the
constant term entering the Friedmann equation reproduces a positive
cosmological constant.

\section{Bosonic matter fields in curved space}\label{sec4}

Given the formulation of gravity described in the previous section, we
shall now extend the constructions presented in  \S~\ref{sec2} to
describe the proper dynamics of the matter fields in curved spacetime. 

\subsection{Scalar field}\label{subsec4.2}

Following \S~\ref{subsec2.2} we assume that the clock field $\phi$ has a derivative coupling to a real scalar field $\chi$ of the form 
\begin{equation}
 \int \dd x^4\sqrt{g_{\rm E}}
  (g_{\rm E}^{\mu\nu}\partial_{\mu}\phi\partial_{\nu}\chi)^2. 
\end{equation}
Adding this to the Riemannian kinetic term and a potential term $\tilde V$ for $\chi$, the general action for $\chi$ is of the form 
\begin{eqnarray}
 S_{\chi}& =& \int \dd x^4 \sqrt{g_{\rm E}}
  \left[
   -\frac{\kappa_{\chi}}{2}
   g_{\rm E}^{\mu\nu}\partial_{\mu}\chi\partial_{\nu}\chi -\tilde{V}(\chi)\right.\nonumber\\
   &&\left.\qquad+ \frac{\alpha_{\chi}}{2M^4}
   (g_{\rm E}^{\mu\nu}\partial_{\mu}\phi\partial_{\nu}\chi)^2 
  \right].
\end{eqnarray}
It involves two dimensionless constants, $\kappa_{\chi}$ and
$\alpha_{\chi}$. There is a freedom to rescale $\chi$ and thus we can
set $\kappa_\chi=\pm1$ if needed. 

With the decomposition (\ref{eqn:EuclideanADM}), the Riemannian kinetic and the derivative coupling terms are correspondingly rewritten as 
\begin{eqnarray}
 g_{\rm E}^{\mu\nu}\partial_{\mu}\chi\partial_{\nu}\chi
  & = & 
  ( \partial^{\rm E}_{\perp}\chi)^2
  + \gamma^{ij}\partial_i\chi\partial_j\chi, 
  \nonumber\\
   (g_{\rm E}^{\mu\nu}\partial_{\mu}\phi\partial_{\nu}\chi)^2 
   & = & 
  \frac{M^4}{N_{\rm E}^2} (\partial^{\rm E}_{\perp}\chi)^2,
\end{eqnarray}
where $\partial^{\rm E}_{\perp}$ is defined in Eq.~(\ref{e.317}). Therefore, the action of the scalar field $\chi$ reduces to
\begin{eqnarray}
 S_{\chi} &=& \int \dd t \dd x^3N_{\rm E}\sqrt{\gamma}
  \left[
 \frac{1}{2}\left(\frac{\alpha_{\chi}}{N_{\rm E}^2}-\kappa_{\chi}\right)
 ( \partial^{\rm E}_{\perp}\chi)^2 -\tilde{V}(\chi)\right.\nonumber\\
&&\qquad\qquad\left.
  -\frac{\kappa_{\chi}}{2}\gamma^{ij}\partial_i\chi\partial_j\chi
  \right].
\end{eqnarray}

If 
\begin{equation}\label{def:parachi}
 \frac{\alpha_{\chi}}{N_{\rm E}^2} > \kappa_{\chi} > 0,
\end{equation}
then $S_{\chi}$ describes a scalar field propagating in a Lorentzian
spacetime. To see this explicitly let us define a Lorentzian effective metric
$g^{\chi}_{\mu\nu}$ by 
\begin{equation}
 g^{\chi}_{\mu\nu}\dd x^{\mu}\dd x^{\nu}
  = -N_{\chi}^2\dd t^2 
  + \Omega_{\chi}^2\gamma_{ij}(\dd x^i+N^i\dd t)(\dd x^j+N^j\dd t), 
\end{equation}
where 
\begin{equation}
 N_{\chi}= N_{\rm E}
  \left[\frac{\kappa_{\chi}^3}
   {\frac{\alpha_{\chi}}{N_{\rm E}^2}-\kappa_{\chi}}\right]^{1/4}, 
  \ 
 \Omega_{\chi} =
  \left[\kappa_{\chi}
   \left(\frac{\alpha_{\chi}}{N_{\rm E}^2}-\kappa_{\chi}\right)\right]^{1/4}. 
\end{equation}
The scalar field action $S_{\chi}$ is rewritten as
\begin{equation}\label{e:S-chi}
 S_{\chi} = -\int \dd x^4\sqrt{-g^{\chi}}
  \left[\frac{1}{2}g_{\chi}^{\mu\nu}
   \partial_{\mu}\chi\partial_{\nu}\chi + V(\chi,X)\right],
\end{equation}
where $g^{\chi}$ and $g_{\chi}^{\mu\nu}$ are the determinant and the
inverse of $g^{\chi}_{\mu\nu}$, and 
\begin{eqnarray}
 V(\chi,X) &= &\tilde{V}(\chi)
  \left[\kappa_{\chi}^3
   \left(\frac{\alpha_{\chi}}{N_{\rm E}^2}-\kappa_{\chi}\right)\right]^{-1/2}
  \nonumber\\
 & = & \tilde{V}(\chi)
  \left[\kappa_{\chi}^3
   \left(\frac{\alpha_{\chi}X_{\rm E}}{M^4}-\kappa_{\chi}\right)\right]^{-1/2}.
\end{eqnarray}
Note that $\alpha_{\chi}$ and $\kappa_{\chi}$ may depend on $X_{\rm E}$ and
that $X_{\rm E}$ is related to $X$ via the correspondence~(\ref{eqn:gravity-correspondence}).

\subsection{Vector field}\label{subsec4.1}

Let us now consider the case of a gauge field $A^\mu$. Similarly as in \S~\ref{subsec2.2}, we can add a coupling to the clock field $\phi$  of the form
\begin{equation}
 \int \dd x^4\sqrt{g_{\rm E}}
  F_{\rm E}^{\mu\rho}F^{\nu}_{{\rm E}\rho}
  \partial_{\mu}\phi\partial_{\nu}\phi
\end{equation}
to the standard (Riemannian) Maxwell action. Again, $F_{\mu\nu}\equiv \partial_{\mu}A_{\nu}-\partial_{\nu}A_{\mu}$ is the Faraday tensor of  $A_{\mu}$ and we use the notations $F^{\mu}_{{\rm E}\nu}\equiv g_{\rm E}^{\mu\rho}F_{\rho\nu}$ and $F_{\rm E}^{\mu\nu}\equiv g_{\rm E}^{\nu\rho}F^{\mu}_{{\rm E}\rho}$.  This leads to the general gauge-invariant  action for the vector field 
\begin{equation}
 S_A = \frac{1}{4}\int dx^4 \sqrt{g_{\rm E}}
  \left[
   -\kappa_A F_{\rm E}^{\mu\nu}F_{\mu\nu} 
   + 2\frac{\alpha_A}{M^4}
   F_{\rm E}^{\mu\rho}F^{\nu}_{{\rm E}\rho}
   \partial_{\mu}\phi\partial_{\nu}\phi 
  \right],
\end{equation}
where $\kappa_A$ and $\alpha_A$ are two dimensionless constants. 

With the decomposition (\ref{eqn:EuclideanADM}) for the Riemannian metric, the Riemannian kinetic term and the non-minimal coupling term can be respectively written as 
\begin{eqnarray}
F_{\rm E}^{\mu\nu}F_{\mu\nu} & = & 
 2\gamma^{ij}\tilde{F}_{\perp i}\tilde{F}_{\perp j}
 +\gamma^{ik}\gamma^{jl}F_{ij}F_{kl},
  \nonumber\\
  F_{\rm E}^{\mu\rho}F^{\nu}_{{\rm E}\rho}
  \partial_{\mu}\phi\partial_{\nu}\phi & = & 
  \frac{M^4}{N_{\rm E}^2}
  \gamma^{ij}\tilde{F}_{\perp i}\tilde{F}_{\perp j},
\end{eqnarray}
where 
\begin{equation}
 \tilde{F}_{\perp i} \equiv 
  \frac{1}{N_{\rm E}}(F_{t i}-N^jF_{ji}).
\end{equation}
Therefore, the gauge-invariant action takes the form
\begin{eqnarray}
 S_A &=& \frac{1}{4}\int \dd t \dd x^3N_{\rm E}\sqrt{\gamma}
  \left[
 2\left(\frac{\alpha_A}{N_{\rm E}^2}-\kappa_A\right)
 \gamma^{ij}\tilde{F}_{\perp i}\tilde{F}_{\perp j}\right.\nonumber\\
&&\qquad\left. -\kappa_A\gamma^{ik}\gamma^{jl}F_{ij}F_{kl}
  \right].
\end{eqnarray}

If 
\begin{equation}\label{def:paraA}
 \frac{\alpha_A}{N_{\rm E}^2} > \kappa_A > 0,
\end{equation}
then $S_A$ describes a $U(1)$ gauge field propagating in a Lorentzian
spacetime. To see this explicitly let us define a Lorentzian effective metric
$g^A_{\mu\nu}$ by 
\begin{equation}
 g^A_{\mu\nu}\dd x^{\mu}\dd x^{\nu}
  = -N_A^2dt^2 
  + \Omega_A^2\gamma_{ij}(\dd x^i+N^i\dd t)(\dd x^j+N^j\dd t), 
\end{equation}
where $\Omega_A$ is an arbitrary positive function and 
\begin{equation}
 N_A = N_{\rm E}\Omega_A
  \left[\frac{\kappa_A}
   {\frac{\alpha_A}{N_{\rm E}^2}-\kappa_A}\right]^{1/2}.  
\end{equation}
The vector field action $S_A$ takes the form of the usual Maxwell action
\begin{equation}\label{e:S-A}
 S_A = -\int \dd x^4\sqrt{-g^A}\frac{1}{4e^2}
  g_A^{\mu\rho}g_A^{\nu\sigma}F_{\mu\nu}F_{\rho\sigma},
\end{equation}
where $g^A$ and $g_A^{\mu\nu}$ are the determinant and the
inverse of $g^A_{\mu\nu}$, and the effective coupling constant $e^2$ is
given by 
\begin{equation}\label{e.U1}
 e^2 = 
  \left[\kappa_A
   \left(\frac{\alpha_A}{N_{\rm E}^2}-\kappa_A\right)
       \right]^{-1/2}.  
\end{equation}
Note that $\alpha_A$ and $\kappa_A$ may depend on $X_{\rm E}$ and 
that $X_{\rm E}$ is related to $X$ via the correspondence
(\ref{eqn:gravity-correspondence}).

\subsection{Generalization to a complex scalar field}

The generalization to a complex scalar field charged under the $U(1)$ is straightforward and follows the construction presented in \S~\ref{subsec2.3}. Consider the action for a complex scalar $\psi$
\begin{eqnarray}
 S_{\psi} & = & \int \dd x^4 \sqrt{g_{\rm E}}
  \left\{
   -\frac{\kappa_{\psi}}{2}g_{\rm E}^{\mu\nu}
   (\partial_{\mu}+iqA_{\mu})\psi^*
   (\partial_{\nu}-iqA_{\nu})\psi
   \right.
   \nonumber\\
 & & \left.
   + \frac{\alpha_{\psi}}{2M^4}
   \left|g_{\rm E}^{\mu\nu}\partial_{\mu}\phi
    (\partial_{\nu}-iqA_{\nu})\psi\right|^2 
   -\tilde{U}(|\psi|^2)
  \right\},
\end{eqnarray}
where $q$, $\kappa_{\psi}$ and $\alpha_{\psi}$ are dimensionless
constants and $\tilde{U}(|\psi|^2)$ is a function of $|\psi|^2$. 

Supposing that
\begin{equation}
 \frac{\alpha_{\psi}}{N_{\rm E}^2} > \kappa_{\psi} > 0,
\end{equation}
it is easy to show that 
\begin{eqnarray}
 S_{\psi} & = & -\int \dd x^4\sqrt{-g^{\psi}}
  \bigg[\frac{1}{2}g_{\psi}^{\mu\nu}
   (\partial_{\mu}+iqA_{\mu})\psi^*
   (\partial_{\nu}-iqA_{\nu})\psi \nonumber\\
 & & 
  \qquad\qquad + U(|\psi|^2,X)\bigg],
\end{eqnarray}
where we have introduced a Lorentzian metric $g^{\psi}_{\mu\nu}$ by
\begin{equation}
 g^{\psi}_{\mu\nu}\dd x^{\mu}\dd x^{\nu}
  = -N_{\psi}^2\dd t^2 
  + \Omega_{\psi}^2\gamma_{ij}(\dd x^i+N^i\dd t)(\dd x^j+N^j\dd t), 
\end{equation}
\begin{equation}
 N_{\psi}= N_{\rm E}
  \left[\frac{\kappa_{\psi}^3}
   {\frac{\alpha_{\psi}}{N_{\rm E}^2}-\kappa_{\psi}}\right]^{1/4}, 
  \ 
 \Omega_{\psi} =
  \left[\kappa_{\psi}
   \left(\frac{\alpha_{\psi}}{N_{\rm E}^2}-\kappa_{\psi}\right)\right]^{1/4},
\end{equation}
$g^{\psi}$ and $g_{\psi}^{\mu\nu}$ are the determinant and the
inverse of $g^{\psi}_{\mu\nu}$, and 
\begin{eqnarray}
 U(|\psi|^2,X) &= &\tilde{U}(|\psi|^2)
  \left[\kappa_{\psi}^3
   \left(\frac{\alpha_{\psi}}{N_{\rm E}^2}-\kappa_{\psi}\right)\right]^{-1/2}
  \\
 & = & \tilde{U}(|\psi|^2)
  \left[\kappa_{\psi}^3
   \left(\frac{\alpha_{\psi}X_{\rm E}}{M^4}-\kappa_{\psi}\right)\right]^{-1/2}.
  \nonumber
\end{eqnarray}
Note that $\alpha_{\psi}$ and $\kappa_{\psi}$ may depend on $X_{\rm E}$
and 
that $X_{\rm E}$ is related to $X$ via the correspondence
(\ref{eqn:gravity-correspondence}).

Generalization to a non-Abelian group is trivial.

\subsection{Massive point particle}

For a massive point particle, we assume that the action is given by
\begin{eqnarray}
 S_{\rm pp} & = &
  \frac{1}{2}\int
  \bigg[{\cal N}^{-1}
 \left(\bar{\kappa}_{\rm pp}g^{\rm E}_{\mu\nu}
  -\frac{\bar{\alpha}_{\rm pp}}{M^4}
  \partial_{\mu}\phi\partial_{\nu}\phi\right)
 \frac{dx^{\mu}}{d\tau}\frac{dx^{\nu}}{d\tau}
 \nonumber\\
 & &  - {\cal N}m^2\bigg] d\tau, 
\end{eqnarray}
where ${\cal N}$ is a function of $\tau$. Then, a test particle propagates in the effective metric
$g^{\rm pp}_{\mu\nu}= \bar{\kappa}_{\rm pp}g_{\mu\nu}^{\rm E} - \frac{\bar{\alpha}_{\rm pp}}{M^4}\partial_{\mu}\phi\partial_{\nu}\phi$ so that its equation of motion is simply a geodesic equation for this metric
\begin{equation}
 u^\mu\nabla_\mu^{\rm pp} u^\nu =0,
\end{equation}
with $u^\mu=\dd x^\mu/\dd\lambda$, where $\lambda$ is an affine 
parameter defined by $d\lambda={\cal N}d\tau$. Using the
decomposition~(\ref{eqn:EuclideanADM}) of the Riemannian metric, we 
obtain that 
\begin{eqnarray}  \label{eqn:EuclideanADM-2}
 g^{\rm pp}_{\mu\nu}\dd x^{\mu}\dd x^{\nu} &=& 
  -\left(\bar{\alpha}_{\rm pp} -\bar{\kappa}_{\rm pp} N_{\rm E}^2\right) 
  \dd t^2 \\
 && +\bar{\kappa}_{\rm pp} \gamma_{ij}
  (\dd x^i+N^i\dd t)(\dd x^j+N^j\dd t).\nonumber 
\end{eqnarray}
This effective metric has Lorentzian signature if 
\begin{equation}
 \frac{\bar{\alpha}_{\rm pp}}{N_{\rm E}^2} > \bar{\kappa}_{\rm pp} > 0.
\end{equation}

\section{Phenomenology}\label{sec5}

\subsection{Summary}

We have proposed a Riemannian field theory for gravity, vector and
scalar fields that, with the expense of the introduction of a scalar field
$\phi$ called a clock field,  leads to an effective Lorentzian dynamics. 

This construction involves a set of free parameters:
\begin{itemize}
 \item For the gravitational sector, we have two free functions of
       $X_{\rm E}$, ${\cal K}$ and $G_4$ in terms of which the two free
       functions of the Lorentzian theory $f(X)$ and $P(X)$ are defined;
       see the correspondence between the two sets given in
       Eq.~(\ref{eqn:gravity-correspondence}). 
 \item For the matter sector, we have derived the actions for scalar and
       vector fields.  Each action depends on 2 parameters
       ($\kappa$, $\alpha$) that are allowed to be functions of 
       $X_{\rm E}$, or equivalently $X$, in general but may as well be
       assumed constant. 
\item Besides, there is an environmental parameter which characterizes
      the clock field configuration on the patch ${\cal M}_0$, 
      $N_{\rm E}$, [see Eqs.~(\ref{e.33}) and~(\ref{def:NE})] and the
      associated integration constant $N_c$ [see
      Eq.~(\ref{e.234})]. They combine in the parameter $r$ [see
      Eq.~(\ref{def:r})]. 
\end{itemize}
With these parameters the actions for gravity, scalar and vector fields
are respectively given by
Eqs.~(\ref{eqn:action-gravity5}),~(\ref{e:S-chi}) and~(\ref{e:S-A}).

We have already shown that these parameters are subject to a series of
constraints.
\begin{itemize}
 \item For the gravitational sector, we have two sets of
       constraints. The first one arises from the stability analysis and
       is given by Eqs.~(\ref{e.tensorC}) and~(\ref{e.scalarC2}). The
       second is related to the dynamics of the homogeneous
       model. Interestingly, the model induces two components which 
       respectively behave as dark matter and dark energy. This sets
       the two constraints~(\ref{e.cosmo1}) and~(\ref{e.cosmo2}) in
       order for the cosmology to be consistent with the standard
       cosmology~\cite{pubook}, at least at the background level. 
 \item For the matter sector, the constants $\alpha$ have to satisfy
       [see Eqs.~(\ref{def:parachi}) and~(\ref{def:paraA})] 
 \begin{equation}\label{e.C1}
  \alpha_\chi >N_{\rm E}^2\kappa_\chi, \qquad
   \alpha_A >N_{\rm E}^2\kappa_A.
 \end{equation}
       From the effective Lagrangians~(\ref{e:S-chi}) and~(\ref{e:S-A}),
       we see that scalars and vectors propagate in 2 different
       effective metrics. In order for the weak equivalence principle to
       hold, we have to impose that these two metrics coincide. This can
       be obtained by imposing that $c_A^2=c_\chi^2$, with 
       $c_A^2=\Omega_A^{-2}N_A^2/N_{\rm E}^2$ and 
       $c_\chi^2=\Omega_\chi^{-2}N_\chi^2/N_{\rm E}^2$. This sets the
       following constraints
\begin{equation}\label{e.C2}
 \frac{\kappa_A}{\alpha_A}=\frac{\kappa_\chi}{\alpha_\chi}.
\end{equation}
       In the simplest situation in which the coefficients
       $(\kappa,\alpha$) are assumed to be constant, we can always set
       $\kappa=1$ in both sectors so that we are left with the
       constraint $\alpha_A=\alpha_\chi$ for the two coupling
       constants. This is similar to what we performed in \S~\ref{sec2}
       in which the couplings to the clock field were chosen a priori so
       that effectively all fields propagate in the same effective
       Minkowski metric. Interestingly, in this class of models, one
       requires a tuning on the parameters of the Lagrangians, but once
       it is done, it is satisfied whatever the configuration of the
       clock field, that is whatever $N_{\rm E}$ or $X_{\rm E}$. In this
       sense the tuning is not worse than the one usually does by assuming
       that all the fields propagate in the same metric. This conclusion 
       holds even if $\kappa$'s and $\alpha$'s are functions of 
       $X_{\rm E}$ as long as their ratios agree between different
       sectors. Again, once this condition is satisfied, it holds
       whatever the field configuration. 

       The Lorentzian effective metric~(\ref{eqn:EuclideanADM-2}) for a
	point particle coincides with that for the vector if 
$\bar{\alpha}_{\rm pp}/\bar{\kappa}_{\rm pp}-N_{\rm E}^2=N_A^2/\Omega_A^2$, 
	that is if 
\begin{equation}
 \frac{\bar{\alpha}_{\rm pp}}
  {\bar{\kappa}_{\rm pp}N_{\rm E}^2}-1 = 
  \left[\frac{\alpha_A}{\kappa_A N_{\rm E}^2}-1\right]^{-1}.
\end{equation}
	This may look as a functional fine-tuning depending on the local
	value of $X_{\rm E}$. Actually, this arises from the fact that
	$\bar{\alpha}_{\rm pp}$ and $\bar{\kappa}_{\rm pp}$ have been
	introduced with reference to $g_{\mu\nu}^{\rm E}$ while
	$\alpha_{A,\chi}$ and $\kappa_{A,\chi}$ have been introduced
	with reference to $g^{\mu\nu}_{\rm E}$. Shifting to the inverse
	metric and redefining these coefficients lead to a constraint
	similar to Eq.~(\ref{e.C2}).

       Only the condition~(\ref{e.C1}) for the emergence of the
       Lorentzian signature is environmentally dependent so that there
       are regions in the configuration space where the dynamics is
       effectively Lorentzian while other regions remain
       Riemannian. This could drive us toward a multiverse description
       in the configuration space but we do not have any anthropic
       reasons associated. 

       Note that without such a tuning so that all fields propagate in
       the same effective Lorentzian metric, one can choose one metric
       as reference so that the equations of motion of other fields will
       exhibit an explicit coupling to the clock field. 
\end{itemize}

\subsection{Other constraints}

The fundamental parameters entering our effective Lorentzian actions are
environmentally determined. This means that if $X_{\rm E}$ is not strictly
constant on ${\cal M}_0$, fundamental constants may be spacetime
dependent, which can induce a violation of the equivalence
principle~\cite{jpucte}. 
\begin{itemize}
 \item From Eq.~(\ref{e.U1}), it can be deduced that the coupling
       constant of any gauge field will be environment-dependent. The
       first implication is that the coupling constants of the three
       non-gravitational interactions have to be space-time
       varying. There exist strong constraints on such a
       possibility~\cite{jpucte}.  
 \item The action for the gravitational sector implies that the Newton
       constant is also expected to be spacetime dependent. 
\item Furthermore, even if $c_A=c_\chi$ so that scalars and vectors
      propagate on the same lightcone, we have to compare the
      propagation speeds of gravity waves and photons. The first is given
      by 
\begin{equation}
 c_\gamma^2=c_A^2 = \frac{N_A^2}{\Omega_A^2N_{\rm E}^2}
  = \left[\frac{\alpha_AX_{\rm E}}{\kappa_A M^4}-1\right]^{-1}.
\end{equation}
      The propagation speed of the gravity waves can be obtained from
      the action~(\ref{e.Tmodes}) rewritten as 
\begin{eqnarray}\label{e.Tmodes2}
 & & \delta S^{(2)}_{{\rm T}, {\bm k}} = \nonumber\\
    & & \frac{1}{8}\int \dd t a^3 N_{\rm E}
  \left[M_{\rm eff}^2\frac{N_{\rm E}}{N} \left(\frac{\dot{h}_{\bm k}}{N_{\rm E}}\right)^2 - 
   2f_0\frac{N}{N_{\rm E}} \frac{{\bm k}^2}{a^2}h_{\bm k}^2 \right],\nonumber
\end{eqnarray}
from which we read off
\begin{equation}
  c^2_{\rm GW} = \frac{2f_0}{M_{\rm eff}^2}\frac{N^2}{N^2_{\rm E}}=\frac{2f_0}{M^2_{\rm eff}}r.
\end{equation}
      It can be rewritten in terms of the function $G_4(X_E)$ entering
      the Euclidean gravitational action (\ref{eqn:action-gravity}) as
\begin{equation}
  c^2_{\rm GW} = \left[\frac{2G'_4X_{\rm E}}{G_4} -1 \right]^{-1}.
\end{equation}
      As long as both lightcones are non-degenrate, there is no a priori
      intrinsic problem even if these two propagation speeds are
      different~\cite{gef,superluminal} and similar features indeed appear in
      many bimetric theories such as TeVeS~\cite{teves} or many other
      extensions of general relativity~\cite{gef}. This difference can
      be tested by future experiments by comparing e.g. the arrival time
      of gravity waves and light emitted during the explosion of
      supernovae; see e.g. Ref.~\cite{woodard}. Models in which 
      $c^2_{\rm GW}<c^2_A$ are very constrained by the observations of
      cosmic rays~\cite{cerenkov} because particles propagating faster
      than the gravity waves emit gravi-Cerenkov radiation. They lead to
      the constraint~\cite{kimura} 
\begin{equation}
 \frac{c_{\gamma}-c_{\rm GW}}{c_{\gamma}}<2\times10^{-15}.
\end{equation}
\end{itemize}

\subsection{Emergence of Lorentz symmetry on intermediate
  scales}\label{subsec5.3} 

Let us first consider our mechanism in flat space. As discussed in
\S~\ref{sec2}, $M$ is not the most important mass scale of the
problem. More important is the scale characterizing the variation of
$X_{\rm E}/M^4$. In order to illustrate this and to capture the way the
Lorentz dynamics appear on relevant scales, let us assume that the clock 
field configuration is given by 
$$
\phi(x^\mu)=M[m_x x+\cos(m_y y)+\beta\cos(M_y y)]
$$ 
with $m_y\ll M_y$ two mass scales that characterize respectively large-
and small-scale variations of $\phi$ and $\beta$ a dimensionless
number. We neglect the two other dimensions for simplicity. With such a
form it is obvious that $\partial_{\mu}\phi$ is not constant. 

Now, assume that we are smoothing the dynamics at a scale 
$R\sim m_x^{-1}$ with e.g. a top-hat window function. On the scale $R$,
the clock field is given by 
\begin{eqnarray}
\phi &=&M\left[m_x x +2\frac{J_1(m_y R)}{m_y R}\cos(m_y y) \right.\nonumber\\
&&\left.\qquad\qquad+2\beta\frac{J_1(M_y R)}{M_y R}\cos(M_y y) 
\right]\nonumber
\end{eqnarray}
where $J_1$ is the Bessel function of order 1. It follows that
$\partial_{\mu}\phi$ is given by 
$$
 \frac{\partial_{\mu}\phi}{M^2}=\left(
\begin{array}{l}
 m_x/M\\
 -2\frac{J_1(m_y R)}{MR}\sin(m_y y) -2\beta\frac{J_1(M_y R)}{MR}\sin(M_y y) 
\end{array} 
 \right).
$$
This can be considered as constant only if $2J_1(m_y R)\ll m_xR$ and 
$2\beta J_1(m_y R) \ll m_xR$. By choosing $R\sim m_x^{-1}$, the first
condition reduces to 
$$
 \frac{m_y}{m_x}\ll 1,
$$
since $J_1(x)\sim x/2$ at small $x$, while the second gives
$$
2\sqrt{\frac{2}{\pi}}\beta\cos\left(\frac{M_y}{m_x}+\frac{\pi}{4}\right)\sqrt{\frac{m_x}{M_y}}\ll1.
$$
This condition is clearly fulfilled if $m_y\ll m_x\ll \beta^{-2}M_y$. 
For example, if we assume that $M_y$ is of the order of the Planck mass,
$M_y\sim M_{\rm p}\sim10^{19}$~GeV and that the large-scale variations
appear on Hubble scales, $m_y\sim H_0\sim 10^{-41}$~GeV, then we end up
with the conclusion that $\partial_{\mu}\phi$ can be considered as
constant at the level $10^{-n}$ on scales 
\begin{equation}
 10^{n-41}\,{\rm GeV} < m_x < 
  10^{21 - 2n}\left(\frac{\beta}{0.1}\right)^{-2}\,{\rm GeV}.
\end{equation}
For e.g. $n=9$ and $\beta={\cal O}(0.1)$, this means that we can work with
scales 
\begin{equation}
 10^{-19}\,{\rm m} < m_x^{-1} < 10^{16}\,{\rm m}.
\end{equation}
\begin{widetext}

\begin{figure}[h!]
\centering
\includegraphics[width=6cm]{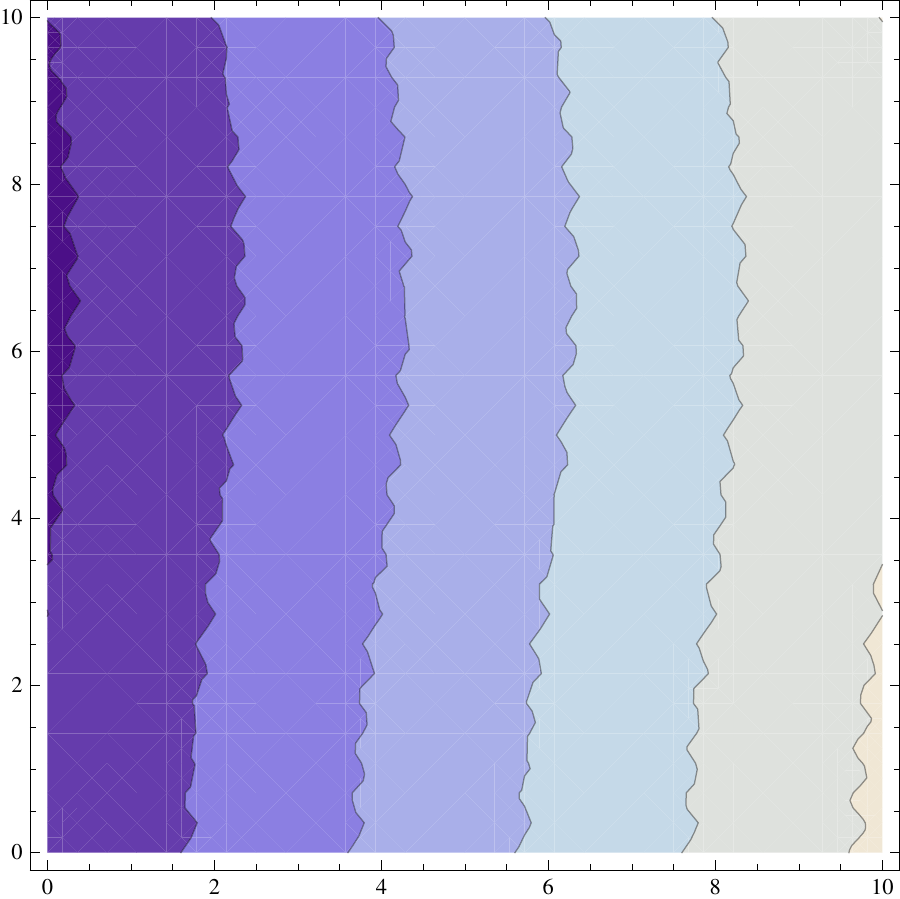}\includegraphics[width=6cm]{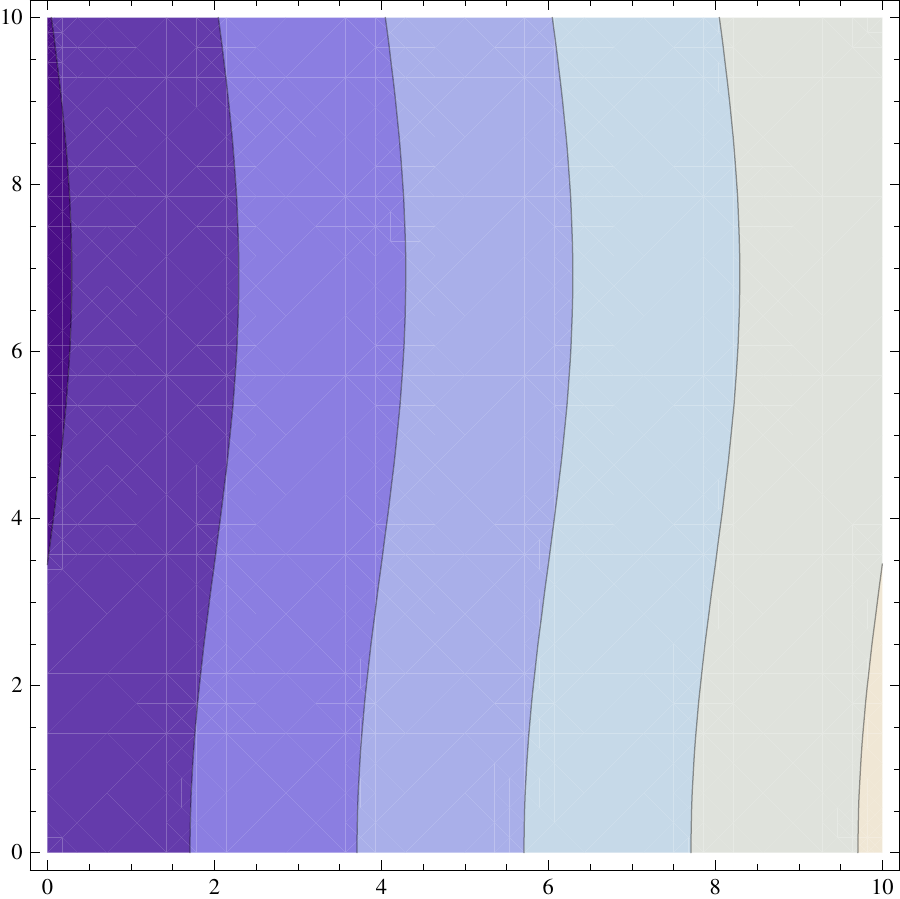}\includegraphics[width=6cm]{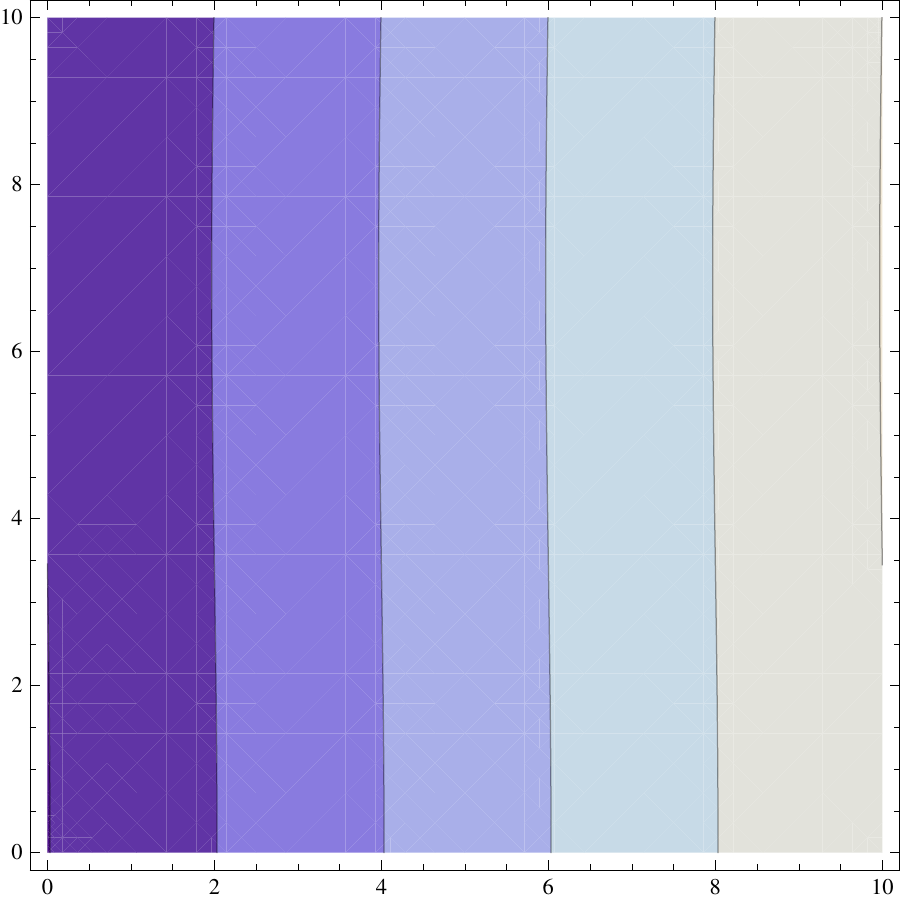}
\caption{Example of a field configuration with fluctuations on scales
 larger than $M_y^{-1}$ and shorter than $m_y^{-1}$ (left). When
 smoothed on a scales $m_x^{-1}$ (middle) and $10m_x^{-1}$ (right) the
 distribution of the clock field is such that $\partial_\mu\phi$ can be
 considered as constant on scales smaller than $m_x^{-1}$.}
\label{fig2}
\end{figure}
\end{widetext}
In such a range of scales, we expect no deviations larger than $10^{-9}$
to the standard field theory. On cosmological scales, we can probably
relax the bound to deviations of order $10^{-1}-10^{-2}$ so that our
model may be compatible with standard cosmology on scales of the order
of the observable universe.

\section{Summary and discussion}\label{sec7}

We have followed the idea that the apparent Lorentzian dynamics of the
usual field theories is an emergent property and that the
underlying field theory is in fact strictly Riemannian. This requires
the introduction of the clock field, a scalar field playing the role of
the physical time. We emphasize that the microscopic theory is
Euclidean, and that time evolution is just an effective and emergent
property, that holds on some energy scales, and in some regions 
of the Euclidean space. We have thus to think of time and dynamics as
illusions in our local patch ${\cal M}_0$. This has to be distinguished 
from the mathematical trick of a Wick rotation used to effectively study
genuine Lorentzian theories in a Euclidean space. 

We have been able to perform such a construction in flat spacetime for
scalar, vector and spinor, hence allowing for the construction of the
standard model of particle physics. Our construction is however
restricted to classical field theory and the spinor sector suffers from
the severe fine-tuning to ensure the CPT invariance. (See
e.g. Ref.~\cite{Toma:2012xa} and references therein for recent
constraints on CPT violation.) 

We have then generalized our construction to curved spacetimes. This
generalizes an early attempt~\cite{Girelli} that ended up with a
Nordstr\"om theory of gravity. Our construction leads to an extended
$K$-essence model for gravity called covariant Galileon, which can be
close enough to general relativity to be experimentally acceptable. We
have then generalized the scalar and vector sectors to curved
spacetimes. This requires the introduction of 4 arbitrary
functions. Again, so far we have not generalized the construction of
spinors to curved spacetime. We have expressed the effective Lorentzian
action that can emerge in a patch ${\cal M}_0$ of the Euclidean
space. It allowed us to list a set of constraints arising from the
stability of the cosmological solution, and the requirement for the 
different test fields to propagate in the same metric, in order for the
weak equivalence principle to hold. The effective fundamental constants,
such as the three non-gravitational coupling constants and the
gravitational constant, are spacetime dependent and the difference of
the propagation speeds of gravity and electromagnetic waves can also set
constraints on our model. We have proposed an heuristic description on
the way the Lorentz symmetry can emerge on a band of energy scales. 

 From a theoretical point of view, our construction gives a new insight
into the need for Lorentzian metric as a fundamental entity. As we have
shown, this is not a mandatory requirement and a decent field theory, at
least at the classical level, can be constructed from a Riemannian
metric. Such a formalism may be fruitful in the debate on the emergence
of time and, speculating, for the development of quantum gravity. 

It also opens up a series of questions and possibilities that will be
addressed in a companion article~\cite{companion}. We can list (1) the
construction of Majorana and Weyl spinors, (2) the development of a
quantum theory and study of particle creation~\cite{particlecreation},
(3) the possibility from the classical viewpoint that singularities in
our local Lorentzian region may be related to singularities in the clock
field (e.g. similar to topological defects) and not in the metric of the
Euclidean theory (see Ref.~\cite{Frolov:2010yj} for a similar idea in a
totally different setup), (4) the possibility that a de Sitter spacetime
may be an ``illusion'' in an anti-de Sitter Riemannian space. It then
follows that a Euclidean AdS/CFT correspondence at the microscopic level
would reveal itself as a dS/CFT correspondence in our effective
Lorentzian universe. 

All these are indeed, for now, bold speculations but they illustrate
that this framework may be fruitful for extending our current field
theories, including general relativity.

\section*{Acknowledgements}
We thank Emir G\"umr\"uk\c{c}\"uo\u{g}lu, John Kehayias, Chunshan Lin,
and Yi Wang for discussions. SM thanks the Institute of Astrophysics for
hospitality during the beginning of this work. JPU thanks the Kavli-IPMU
in Tokyo and the Yukawa Institute for Theoretical Physics at Kyoto
University,  where this work was completed during the Long-term Workshop
YITP-T-12-03 on ``Gravity and Cosmology 2012''.  The work of SM was
supported by Grant-in-Aid for Scientific Research 24540256 and 21111006,
by Japan-Russia Research Cooperative Program and by the WPI Initiative,
MEXT, Japan. JPU acknowledges partial support from the Agence Nationale
de la Recherche via the Grant THALES (ANR-10-BLAN-0507-01-02) and thanks
St\'ephane Charlot and the PNCG for their support. 



\begin{references}

 
\bibitem{topology}
 J.-P. Luminet, \etal, ``Dodecahedral space topology as an explanation for weak wide-angle temperature correlations in the cosmic microwave background'',  Nature (London) {\bf425}, 593 (2003), [{\tt astro-ph/0310253}].
  
\bibitem{jpucte}
 J.-P. Uzan,
 ``The fundamental constants and their variation: observational status and theoretical motivations'',
 Rev. Mod. Phys. {\bf 75}, 403 (2003), [{\tt hep-ph/0205340}];
 J.-P. Uzan,
 ``Variation of the constants in the late and early universe'',
 AIP Conf. Proc. {\bf736}, 3 (2004), [{\tt astro-ph/0409424}];
 J.-P. Uzan,
 ``Varying constants, Gravitation and Cosmology'',
 Living Rev. Relat. {\bf4}, 2 (2011), [{\tt arXiv:1009.5514}].
 
\bibitem{stringT}
 J. Polchinsky, {\it String theory} (Cambridge University Press, 1998). 

\bibitem{V4}
 I. Bars, ``Survey of two time physics'',
 Class. Quant. Grav. {\bf18}, 3113 (2001), [{\tt }hep-th/0008164];
 G B Halsted, ``Four-Fold Space and Two-Fold Time'', Science {\bf19}, 319 (1892).

\bibitem{V5}
 C.M. Hull, ``Duality and the signature of space-time'',
 JHEP {\bf9811}, 017 (1998), [{\tt hep-th/9807127}]. 
 
 \bibitem{no2}
  J. Dorling, ``The Dimensionality of Time'', Am. J. Phys. {\bf38}, 539 (1970). 
 
\bibitem{gibbons}
 G.W. Gibbons, in Fundamental Theories of Physics, {\bf 164}, ``The Arrows of Time: A debate in Cosmology'' eds L.Mersini-Houghton and R. Vass, pp 107-146 Springer 2012, [{\tt arXiv:1111.0457}].
 
\bibitem{nothing}
 J.B. Hartle, and S.W. Hawking,  ``Wave function of the Universe'', Phys. Rev. D {\bf28}, 2960 (1983);
  J. L. Friedman, ``Lorentzian universes from nothing'', Class. Quant. Grav. {\bf15}, 2639 (1998);
 J.R. Gott, J.R., and X.I. Li, ``Can the universe create itself?'', Phys. Rev. D {\bf58}, 023501 (1998), [{\tt arXiv:astro-ph/9712344}];
 G.W. Gibbons, and J.B. Hartle, ``Real Tunneling Geometries and the Large Scale Topology of the Universe'', Phys. Rev. D {\bf42}, 2458 (1990).

\bibitem{G85}
 A.S. Eddington, {\it The Mathematical Theory of Relativity} (Cambridge University Press, 1923), p. 25.

\bibitem{P86}
 G. W. Gibbons, and A. Ishibashi, ``Topology and signature changes in braneworlds'', Class. Quant. Grav. {\bf 21}, 2919 (2004), [{\tt arXiv:hep-th/0402024}];
 M. Mars, J.M.M. Senovilla, and R. Vera, ``Lorentzian and signature chang- ing branes'', Phys. Rev. D {\bf76}, 044029 (2007), [{\tt arXiv:0705.3380 [hep-th]}];
 M. Mars, J.M.M. Senovilla, and R. Vera, ``Is the accelerated expansion evidence of a forthcoming change of signature?'', 
 [{\tt arXiv:0710.0820 [gr-qc]}].
  
\bibitem{lqc} 
  J.~Mielczarek, ``Signature change in loop quantum cosmology'',
  [{\tt arXiv:1207.4657 [gr-qc]}].  
  
\bibitem{zeh}
 H.D. Zeh, 
 {\it The physical basis of the direction of time} 
 (Springer-Verlag, Berlin, 2001).

\bibitem{vuibert}
 N. Deruelle, and J.-P. Uzan, {\it M\'ecanique et gravitation newtonienne} (Vuibert, Paris, 2006).

\bibitem{barbour}
J. Barbour, {\it The End of Time: The Next Revolution In Physics} (Oxford University Press, 2000).

\bibitem{time-RG}
 T. Damour, ``Time and Relativity'', S\'eminaire Poincar\'e XV (2010).
 
\bibitem{rovelli0}
  C. Rovelli, ``Quantum mechanics without time: a model'', Phys. Rev. D {\bf42}, 2638 (1991);
  R. Giannitrapani, ``On a Time Observable in Quantum Mechanics'', [{\tt arXiv:quant-ph/9611015}].
    
 \bibitem{rovelli}
C. Rovelli, ``Forget time'', [{\tt arXiv:0903.3832}].

\bibitem{time}
 V. Balasubramanian, ``What we don't know about time'', [{\tt arXiv:1107.2897}]. 

\bibitem{Ellis:2006sq} 
  G.~F.~R.~Ellis,
  ``Physics in the real universe: Time and spacetime,''
  Gen.\ Rel.\ Grav.\  {\bf 38}, 1797 (2006)
  [gr-qc/0605049].

\bibitem{rovelli2}
  C. Rovelli, ``Quantum evolving constants'', Phys. Rev. D {\bf44}, 1339 (1991);
  C Rovelli,  {\it Quantum Gravity} (Cambridge University Press, 2004);
  C. Rovelli, ``Time in quantum gravity: An hypothesis'', Phys. Rev. {\bf43}, 442-456 (1991).
  C. Rovelli, ``What is observable in classical and quantum gravity?'', Class. Quant. Grav. {\bf8}, 297 (1991);
   W. G. Unruh and R. M. Wald, ``Time and the interpretation of canonical quantum gravity'', Phys. Rev. D {\bf40}, 2598 (1989) 2598.
 
\bibitem{markopoulou}
F. Markopoulou, ``Space does not exist, so time can'', [{\tt arXiv:0909.1861[gr-qc]}];

\bibitem{time-emergence}
P. Martinetti, ``Emergence of time in quantum gravity: is time necessarily flowing?'', [{\tt arxiv:1203.4995}].
 
\bibitem{time-emergence2}
 C.J.Isham, and J. Butterfield, ``On the Emergence of Time in Quantum Gravity'', in {\it The Arguments of Time}, ed. J. Butterfield (Oxford University Press, 1999), [{\tt gr-qc/9901024}].

\bibitem{time-emergence3}
 M. Heller, and W. Sasin, ``Emergence of Time'', Phys. Lett. A {\bf250}, 48 (1998), [{\tt gr-qc/9711051}]

\bibitem{time-emergence0}
 A. Connes, and C. Rovelli, ``Von Neumann Algebra Automorphisms and Time-Thermodynamics Relation in General Covariant Quantum Theories'', Class. Quant. Grav. {\bf11}, 2899 (1994), [{\tt gr-qc/9406019}].

 \bibitem{carroll}
S. M. Carroll, ``What if Time Really Exists?'', [{\tt arXiv:0811.3772 [gr- qc]}];
G.F.R. Ellis, and T. Rothman, ``Time and Spacetime: The Crystallizing Block Universe'', Int. J. Theor. Phys. {\bf49}, 988 (2010), [{\tt arXiv:0912.0808}];
G.F. R. Ellis, and R. Goswami, ``Space time and the passage of time'', [{\tt arXiv:1208.2611}].

\bibitem{maldacena}
 J. M. Maldacena, ``The Large N Limit of Superconformal Field Theories and Supergravity'', Adv. Theor. Math. Phys. {\bf2}, 231 (1998), [{\tt arXiv:hep-th/9711200}].

\bibitem{holo}
 G. 't Hooft, ``Dimensional Reduction in Quantum Gravity'', {[\tt arXiv:gr-qc/9310026}],
 L. Susskind, ``The World as a Hologram'', J. Math. Phys. {\bf36}, 6377 (1995) [{\tt arXiv:hep-th/9409089}].

\bibitem{V6}
 V. Balasubramanian, \etal, ``Quantum geometry and gravitational entropy'',
 JHEP {\bf 0712}, 067 (2007), [{\tt arXiv:0705.4431}];
 T. Jacobson, ``Thermodynamics of space-time': the Einstein equation of state'';
 Phys. Rev. Lett. {\bf75}, 1260 (1995), [{\tt gr-qc/9504004}];
 E.P. Verlinde, ``On the origin of gravity and the laws of Newton'',
 JHEP {\bf 1104}, 029 (2011), [{\tt arXiv:1001.0785}];
 T. Padmanabhan, ``Thermodynamical Aspects of Gravity: New insights'', Rep. Prog. Phys. {\bf73}, 046901 (2010),
 [{\tt arXiv:0911.5004}].

\bibitem{Greensite:1992np} 
  J.~Greensite,
  ``Dynamical origin of the Lorentzian signature of space-time,''
  Phys.\ Lett.\ B {\bf 300}, 34 (1993)
  [gr-qc/9210008].

\bibitem{Girelli}
F. Girelli, S. Liberati, and L. Sindoni, 
``On the emergence of Lorentzian signature and scalar gravity'',
Phys. Rev. D {\bf79}, 044019 (2009), [{\tt arXiv:0806.4239}].

\bibitem{Goulart}
  E.~Goulart and S.E.P.~Bergliaffa,
  ``Effective metric in nonlinear scalar field theories'',
  Phys.\ Rev. D {\bf 84}, 105027 (2011), [{\tt arXiv:1108.3237}].

\bibitem{Wetterich:2010ni} 
  C.~Wetterich,
  ``Spinors in Euclidean field theory, complex structures and discrete symmetries,''
  Nucl.\ Phys.\ B {\bf 852}, 174 (2011)
  [{\tt arXiv:1002.3556 [hep-th]}].

\bibitem{mehta}
 M.R. Metha,
 ``Euclidean continuation of the Dirac fermion'',
 Phys. Rev. Lett. {\bf 65}, 1983 (1990).

\bibitem{Horndeski:1974wa} 
  G.W.~Horndeski,
  ``Second-order scalar-tensor field equations in a four-dimensional space'',
  Int.\ J.\ Theor.\ Phys.\  {\bf 10}, 363 (1974).

\bibitem{Deffayet:2011gz} 
  C. Deffayet, G. Esposito-Farese, A. Vikman, 
  ``Covariant Galileon'',
  Phys.Rev.D {\bf79}, 084003 (2009), [{\tt arXiv:0901.1314}];
  C.~Deffayet, X.~Gao, D.A.~Steer and G.~Zahariade,
  ``From k-essence to generalised Galileons'',
  Phys.\ Rev.\ D {\bf 84}, 064039 (2011), [{\tt arXiv:1103.3260}].

\bibitem{Nicolis:2008in} 
  A.~Nicolis, R.~Rattazzi and E.~Trincherini,
  ``The Galileon as a local modification of gravity,''
  Phys.\ Rev.\ D {\bf 79}, 064036 (2009)
  [{\tt arXiv:0811.2197 [hep-th]}].

\bibitem{christos} 
  C.~Charmousis, E.J.~Copeland, A.~Padilla and P.M.~Saffin,
  ``General second order scalar-tensor theory, self tuning, and the Fab Four'',
  Phys.\ Rev.\ Lett.\  {\bf 108}, 051101 (2012), [{\tt arXiv:1106.2000}].  
  
\bibitem{ArkaniHamed:2003uy}
  N.~Arkani-Hamed, H.-C.~Cheng, M.A.~Luty and S.~Mukohyama,
  ``Ghost Condensation and a Consistent Infrared Modification of Gravity'',
  JHEP {\bf 0405}, 074 (2004), [{\tt hep-th/0312099}].

\bibitem{ArkaniHamed:2005gu}
  N.~Arkani-Hamed, H.-C.~Cheng, M.A.~Luty, S.~Mukohyama and T.~Wiseman,
  ``Ghost Condensation and a Consistent Infrared Modification of Gravity'',
  JHEP {\bf 0701}, 036 (2007), [{\tt hep-ph/0507120}].

\bibitem{Kobayashi:2011nu} 
  T.~Kobayashi, M.~Yamaguchi and J.'i.~Yokoyama,
  ``Generalized G-inflation: Inflation with the most general second-order field equations'',
  Prog.\ Theor.\ Phys.\  {\bf 126}, 511 (2011), [{\tt arXiv:1105.5723 [hep-th]}].

\bibitem{pubook}
 P. Peter and J.-P. Uzan, 
 {\it Primordial cosmology} 
 (Oxford University Press, England, 2009).

\bibitem{gef}
 J.-P. Bruneton, and G. Esposito-Farese,
 ``Field-theoretical formulations of MOND-like gravity'',
  Phys. Rev. D {\bf76}, 124012 (2007), [{\tt arXiv:0705.4043}];

\bibitem{superluminal}
 J.-P. Bruneton, ``Causality and Superluminal Fields'', [{\tt arXiv:hep-th/0612113}];
 E. Babichev, V. Mukhanov, and A. Vikman, ``k-Essence, superluminal propagation, causality and emergent geometry'', JHEP {\bf0802}, 101 (2008), [{\tt arXiv:0708.0561}].;
 R. Geroch, ``Faster Than Light?'', [{\tt arXiv:1005.1614}];
 G.F.R. Ellis, and J.-P. Uzan, ``$c$ is the speed of light, isn't it?'' Am. J. Phys. 73 (2005) 240, [{\tt gr-qc/0305099}]. 

\bibitem{teves}
 J. Bekenstein, ``Relativistic gravitation theory for the MOND paradigm'', Phys. Rev. D {\bf70}, 083509 (2004), [{\tt astro-ph/0403694}].

\bibitem{woodard}
 E.O. Kahya, and R.P. Woodard,  ``A Generic Test of Modified Gravity Models which Emulate Dark Matter'', 
 Phys. Lett. B {\bf652}, 213 (2007), [{\tt arXiv:0705.0153}]. 
  
\bibitem{cerenkov}
 C.M. Caves, Ann. Phys. {\bf125}, 35 (1980);
 G.D. Moore and A.E. Nelson, ``Lower Bound on the Propagation Speed of Gravity from Gravitational Cherenkov Radiation'', JHEP {\bf09}, 023 (2001), [{\tt hep-ph/0106220}];
 J.W. Elliott, G.D. Moore, and H. Stoica, ``Constraining the New Aether: Gravitational Cherenkov Radiation'', JHEP {\bf08}, 066 (2005), [{\tt hep-ph/0505211}].  

\bibitem{kimura}
 R. Kimura, and K. Yamamoto, ``Constraints on general second-order scalar-tensor models from gravitational Cherenkov radiation'', JCAP {\bf1207}, 050 (2012), [{\tt arXiv:1112.4284}].
 
\bibitem{Toma:2012xa} 
 K.~Toma, S.~Mukohyama, D.~Yonetoku, T.~Murakami, S.~Gunji and T.~Mihara
 {\it et al.}, 
 ``Strict Limit on CPT Violation from Polarization of Gamma-Ray Burst,''
 Phys.\ Rev.\ Lett.\ {\bf 109}, 241104 (2012)
 [arXiv:1208.5288 [astro-ph.HE]].

\bibitem{companion}
Work in progress. 

\bibitem{particlecreation} 
  A.~White, S.~Weinfurtner and M.~Visser,
  ``Signature change events: A Challenge for quantum gravity?,''
  Class.\ Quant.\ Grav.\  {\bf 27}, 045007 (2010)
  [arXiv:0812.3744 [gr-qc]]; 
  S.~Weinfurtner, A.~White and M.~Visser,
  ``Trans-Planckian physics and signature change events in Bose gas hydrodynamics,''
  Phys.\ Rev.\ D {\bf 76}, 124008 (2007)
  [gr-qc/0703117 [GR-QC]].

\bibitem{Frolov:2010yj} 
  V.~P.~Frolov and S.~Mukohyama,
  ``Brane Holes,''
  Phys.\ Rev.\ D {\bf 83}, 044052 (2011)
  [arXiv:1012.4541 [hep-th]].
\end{references}
\end{document}